\documentclass[pre,twocolumn,showpacs,amsmath,amssymb,aps,floatfix,unsortedaddress]{revtex4-1} 
\usepackage{graphicx} 
\usepackage{hyperref} 
\usepackage{xcolor}

\begin{document} 

\graphicspath{{figures/}}

\title{Vibrated polar disks: spontaneous motion, binary collisions, and collective
dynamics}  
\date{\today}

\author{Julien Deseigne} 
\affiliation{LPMCN, Universit\'e de Lyon 1 and CNRS UMR 5586, 69622 Villeurbanne, France} 

\author{S\'ebastien L\'eonard} 
\affiliation{Service de Physique de l'Etat Condens\'e, CEA-Saclay, 91191
Gif-sur-Yvette, France}

\author{Olivier Dauchot}
\affiliation{EC2M-Gulliver, ESPCI-ParisTech and CNRS UMR 7083, 75005 Paris, France}

\author{Hugues Chat\'e} 
\affiliation{Service de Physique de l'Etat Condens\'e, CEA-Saclay, 91191
Gif-sur-Yvette, France}

\begin{abstract}
We study the spontaneous motion, binary collisions, and collective dynamics of ``polar disks'', {\it i.e.} purpose-built particles which, when vibrated between two horizontal plates, move coherently along a direction strongly correlated to their intrinsic polarity.  
The motion of our particles, although nominally three-dimensional and complicated, is well accounted for by
a two-dimensional persistent random walk. Their binary collisions are spatiotemporally extended events during which multiple actual collisions happen, yielding a weak average effective alignment.
We show that this well-controlled, ``dry active matter''  system can display collective motion with orientationally-ordered regions of the order of the system size. We provide evidence of strong number density in the most ordered regimes observed. 
These results are discussed in the light of the limitations of our  system, notably those due to the inevitable presence of walls.
\end{abstract}


\maketitle

\section{Introduction}
\label{intro}

Active matter is nowadays the expression designating out-of-equilibrium systems where energy is spent locally to produce directed motion \cite{ramaswamy2010mechanics}. 
Examples abound, at all scales and in many fields, with perhaps the most spectacular ones in living systems,  from the motion of subcellular components to that of large animal groups \cite{ndlec1997self,kierfeld2008active,ziebert2009collective,schaller467polar,dombrowski2004self,zhang2010collective,ballerini2008interaction,makris2006fish,buhl2006disorder}. 
Active matter is of course the word used by physicists, who have been flocking to this rather new, burgeoning field, producing novel theoretical ideas and results, and creating well-controlled objects that can move ``on their own'' either swimming in a fluid or crawling on a substrate\cite{golestanian2005propulsion,howse2007self,hanczyc2007fatty,palacci2010sedimentation,dos1995free,paxton2004catalytic,lazar2005reversible}. 
Although the fluid is always there, it can sometimes be safely neglected, and momentum conservation thus disobeyed, in particular when objects are moving on a substrate, which acts as a momentum sinks.
One then speaks of ``dry'' active matter \cite{ramaswamy2010mechanics,toner2005hydrodynamics}. 

Microscopic models of dry active matter usually consist
of self-propelled particles whose local interactions give rise 
to some kind of alignment.
This includes the now famous Vicsek {\it et al.} model in which constant-speed
point particles align ``ferromagnetically'' with neighboring ones, in the presence of
noise, in what amounts to an out-of-equilibrium version of the XY model 
\cite{vicsek1995novel}.
Important results have been obtained recently both on microscopic models like the 
Vicsek model and on continuous descriptions of dry active matter
\cite{gregoire2004onset,chate2006simple,chate2008collective,ginelli2009large,bertin2006boltzmann,baskaran2008enhanced,ihle2011kinetic}. 
These include
the seminal calculation by Toner and Tu \cite{toner1998flocks,toner2009fast} 
who confirmed, via a renormalization-group approach, the numerical findings of Vicsek
{\it et al.} that collective motion can be sustained at all scales in spite of 
the presence of noise: true long-range orientational order exists 
even in two dimensions when spins are moving in space.

Although much remains to be done, progress has been recorded in matching
the many numerical results obtained with microscopic models to their
continuous descriptions \cite{bertin2006boltzmann,baskaran2008enhanced,baskaran2008hydrodynamics,ihle2011kinetic}, 
and important theoretical predictions have been
confirmed in numerical studies \cite{chate2006simple,chate2008collective,ginelli2009large}. 
The situation is much less satisfactory at the
experimental level. This should not be a surprise: living systems, be they
large animal groups or collections of cells or even biofilaments and motor proteins,
are difficult if not impossible to control, and they typically involve
many extra, unwanted, and often unknown factors and mechanisms.

\begin{figure*}[!t] 
\hbox{
\hspace{-4.5cm}
\vbox{
\includegraphics[width=9cm]{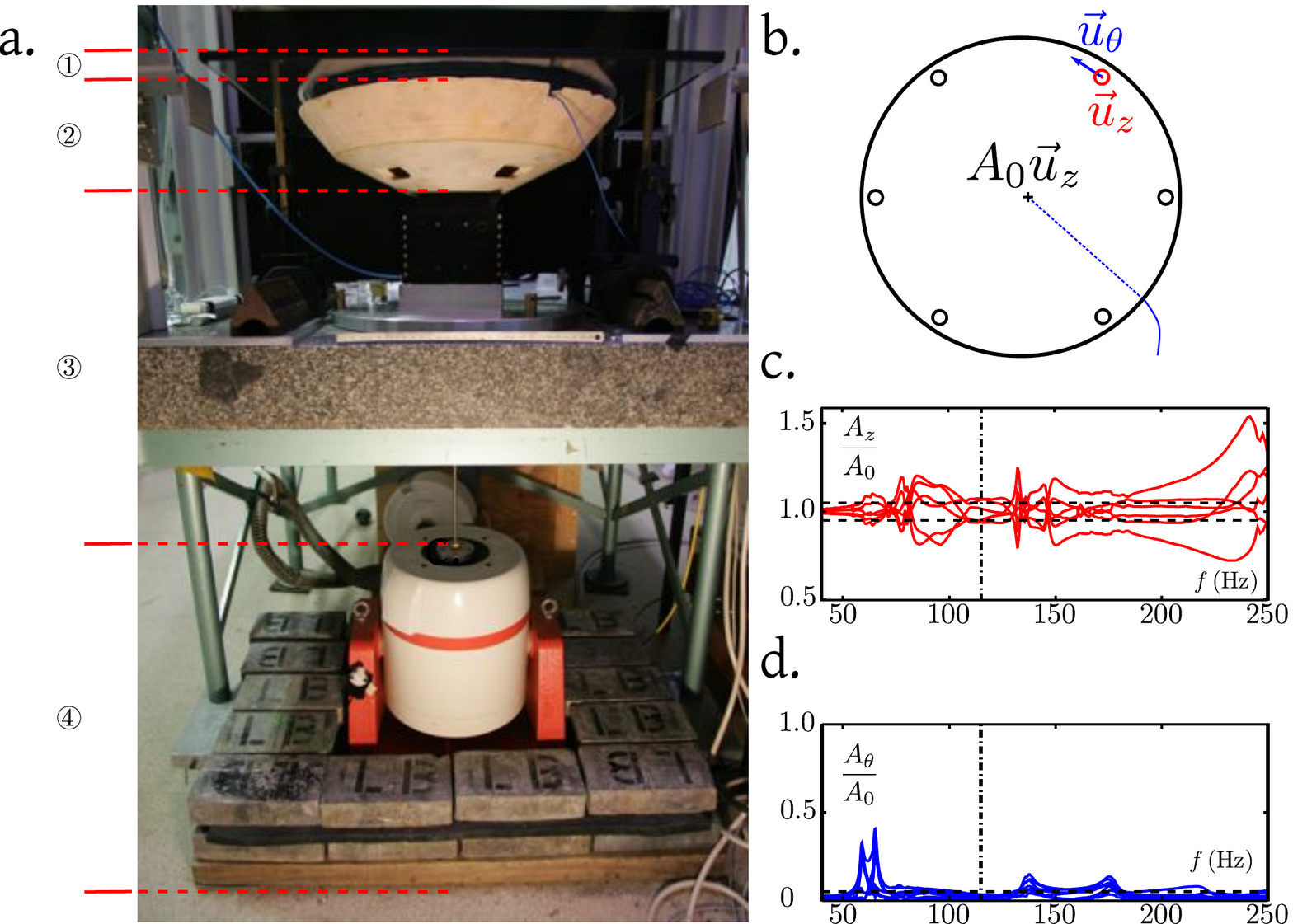}\\
\vspace{2mm}
\hspace{1.5cm}(a)
}
\hspace{-9cm}
\vbox{
\includegraphics[width=7.5cm]{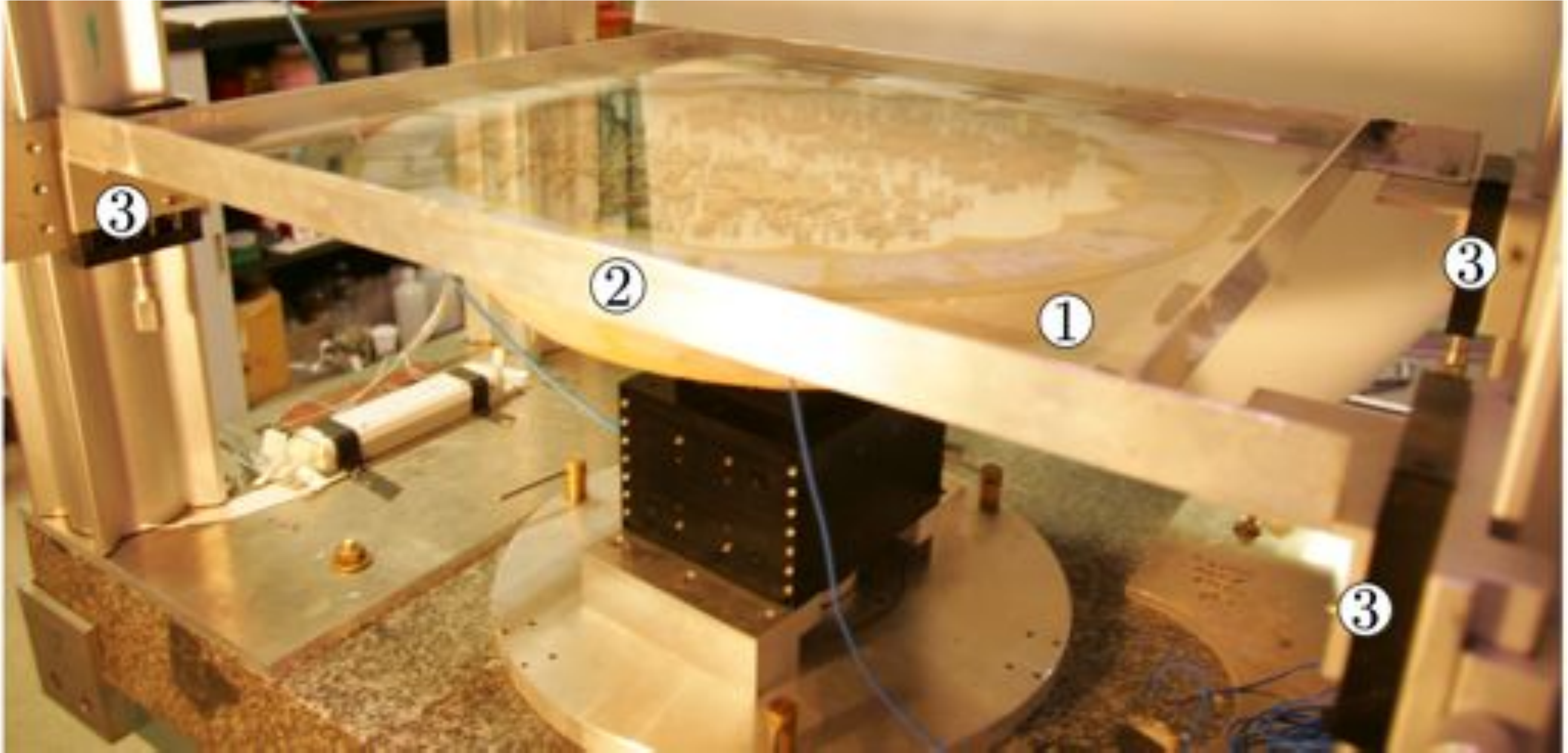}\\
\vspace{2mm}
(b)\\
\vspace{2mm}
\includegraphics[width=8cm]{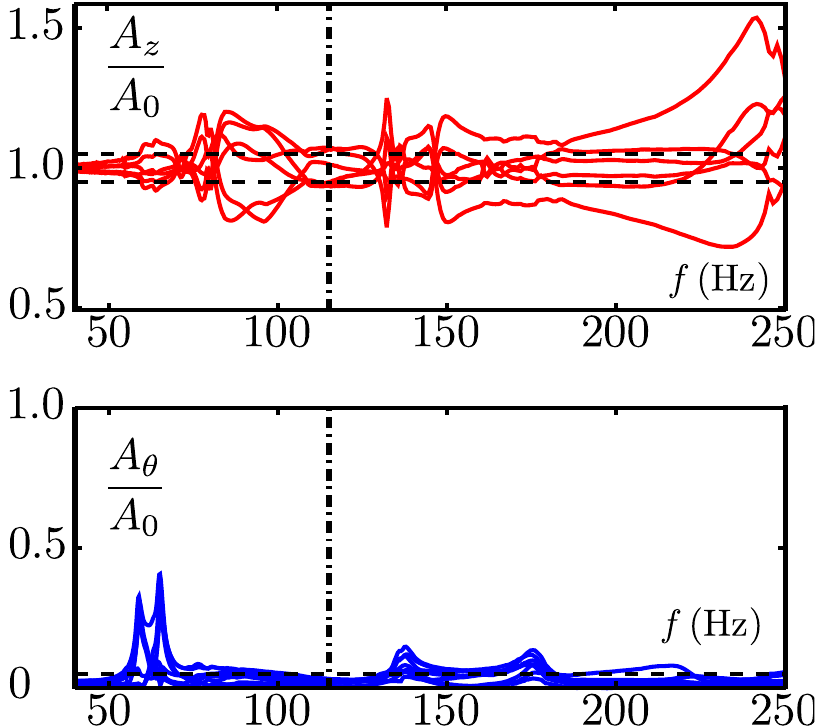}\\
\vspace{2mm}
(c)
}
}
\caption{(color online) Vibrational system. 
(a) The vibration is provided by an electromagnetic servo-controlled shaker ballasted with lead and isolated from ground by rubber mats (4). It is then transmitted to the vibrating plate (2) via a brass rod coupled to a square air bearing (3). The vibrating plate (2) is composed of a nylon plate, an expanded polystyrene truncated cone, on top of which is glued the glass plate which supports the vibrated grains. Finally, the grains are confined vertically by a top glass plate fixed to an external rigid frame (1). 
(b) This top glass plate (1) is in a rigid frame (2), the vertical position of which is finely tuned by micro-control screws (3) to ensure an homogeneous gap of 2.4 mm.
(c) The vibration is imposed and retro-controlled at the center of the glass plate supporting the grains. It is measured at the periphery of the plate. Vertical  and azimuthal transfer functions between the acceleration imposed by the controller in the center of the plate and the one measured at six points on the plate perimeter. The horizontal dashed dark lines indicate a control of the acceleration within $5\%$. The vertical dot-dashed line indicates the frequency at which all experiments reported in this paper have been conducted
}
\label{fig:set-up}
\end{figure*}

In this setting, building inert objects capable of moving by themselves or by
capturing energy from their environment is a crucial step towards well-controlled 
experiments. In the context of dry active matter, two options are available: 
moving ``robots'' interacting by direct contact or at some distance
via sensors \cite{gross2008evolution}, 
or asymmetric objects vibrated on some homogeneous substrate.
Here we follow this last option, building on knowledge gathered over the last
few decades on shaken granular media.

Yamada, Hondou, and Sano were pioneers in demonstrating that an axisymmetric
polar object vibrated between two plates can move quasiballistically
\cite{yamada2003coherent}. Kudrolli's group studied the behavior of polar rods \cite{kudrolli2008swarming}
and, more recently, of short snakelike chains \cite{kudrolli2010concentration}. 
Shaken elongated apolar particles (a realization of so-called active nematics)
have been considered in \cite{narayan2007long}.

The particles above are all elongated objects, with their asymmetry
inscribed in their shape. This usually ensures that
their (inelastic) collisions result in alignment, in a way after all 
rather similar to what happens in numerical models 
\cite{grossman2008emergence,aranson2007swirling,peruani2006nonequilibrium,elgeti2009self}. 
In this paper, we deliberately depart from this: 
 we report further and at length
on our experiments performed on {\it vibrated polar disks},
following a first account of their collective properties \cite{deseigne2010collective}.
The asymmetry of our particles is thus {\it not} in their shape, but in their contact properties with the vibrating plate
(both friction and restitution coefficient), which endows them, as we shall see, with rather unusual collision dynamics.

The rest of this paper is organized as follows: 
in Section~\ref{setup} we detail our vibration apparatus and the design of 
our polar disks.  Section~\ref{single} is devoted to the description of 
the individual motion of our particles and how it varies with our main
control parameter, the vibration amplitude. In  Section~\ref{binary},
we show that collisions between two of our polar disks possess
rather complicated properties, but they do result in some 
effective alignment. In Section~\ref{collective}, we turn to the collective 
properties of our particles and show that our system can produce orientational
order over large scales. Section~\ref{discuss} contains a discussion of 
the limitations of our results and offers some perspectives on future work.

\section{Experimental apparatus}
\label{setup}

In principle, producing some kind of self-propulsion using vibrating 
man-made objects allows for incomparably more control than with, say, 
 biological organisms or even motility assays in which biofilaments are displaced
by motor proteins grafted to a substrate.
Nevertheless, in order to be free from various artifacts,
these experiments require extremely homogeneous and strictly horizontal vibration,
which is notoriously difficult to achieve. Indeed, previous attempts 
have been suspected of being plagued by large-scale spurious modes
\cite{narayan2007long,aranson2008comment,narayan2008response}.
In this section, we first introduce the vibrational system we used 
to generate a well-controlled vertical vibration.
We then describe and motivate the design of our polar disks.

\subsection{Vibration system} 

The vibration is provided by an electromagnetic servo-controlled shaker 
(V455/6-PA1000L,LDS) [Fig.\ref{fig:set-up}a.], which rests on a thick 
wooden plate ballasted with 300\,kg lead bricks and isolated from 
ground with rubber mats (MUSTshock 100x100xEP5).  The vibration is 
transmitted to the vibrating plate via a 400\,mm-long, 8\,mm-thick brad ross, 
coupled with a stiff square air-bearing slider of size 127\,mm 
(C40-03100-100254, IBSPE). The brass rod is flexible enough to compensate 
for the alignment mismatch, but stiff enough to ensure the mechanical coupling 
between the slider and the shaker.  The slider ensures virtually friction-free 
and submicron amplitude horizontal motion. The plate is composed of a 50\,mm 
thick nylon plate fixed to the slide, a  50\,mm expanded polystyrene truncated 
cone, on top of which is glued the 425\,mm diameter glass plate supporting the vibrated 
grains. Finally, the vertical motion of the grains is confined by a top glass 
plate attached to an external non vibrating rigid frame. The two glass plates 
are separated by a gap $h=2.4$\,mm.

An uniaxial accelerometer, used to control the vibration, is glued at 
the bottom of the bottom glass plate, inside a small cavity encarved in 
the  polystyrene. In order to control the vibration quality, we measure with a triaxial 
accelerometer (356B18, PCB electronics) the vertical and azimuthal 
transfer functions between the controller and the acceleration recorded 
on the perimeter of the bottom glass plate, when vibrated at an acceleration 
$A_0 = 1$\,g, where g stands for the acceleration of gravity 
[Fig.\ref{fig:set-up}b.]. 
A working frequency $f = 115$\,Hz was chosen so that it avoids all
resonances in the vibrating system while inducing motion of our polar disks (see below).
At this frequency, the azimuthal to vertical 
acceleration ratio is smaller than $1\%$ and the relative vibration heterogeneities are less than $5\%$.

\subsection{Self-propelled particles}

\begin{figure}[!t]
\centering 
\includegraphics[width=8.4cm]{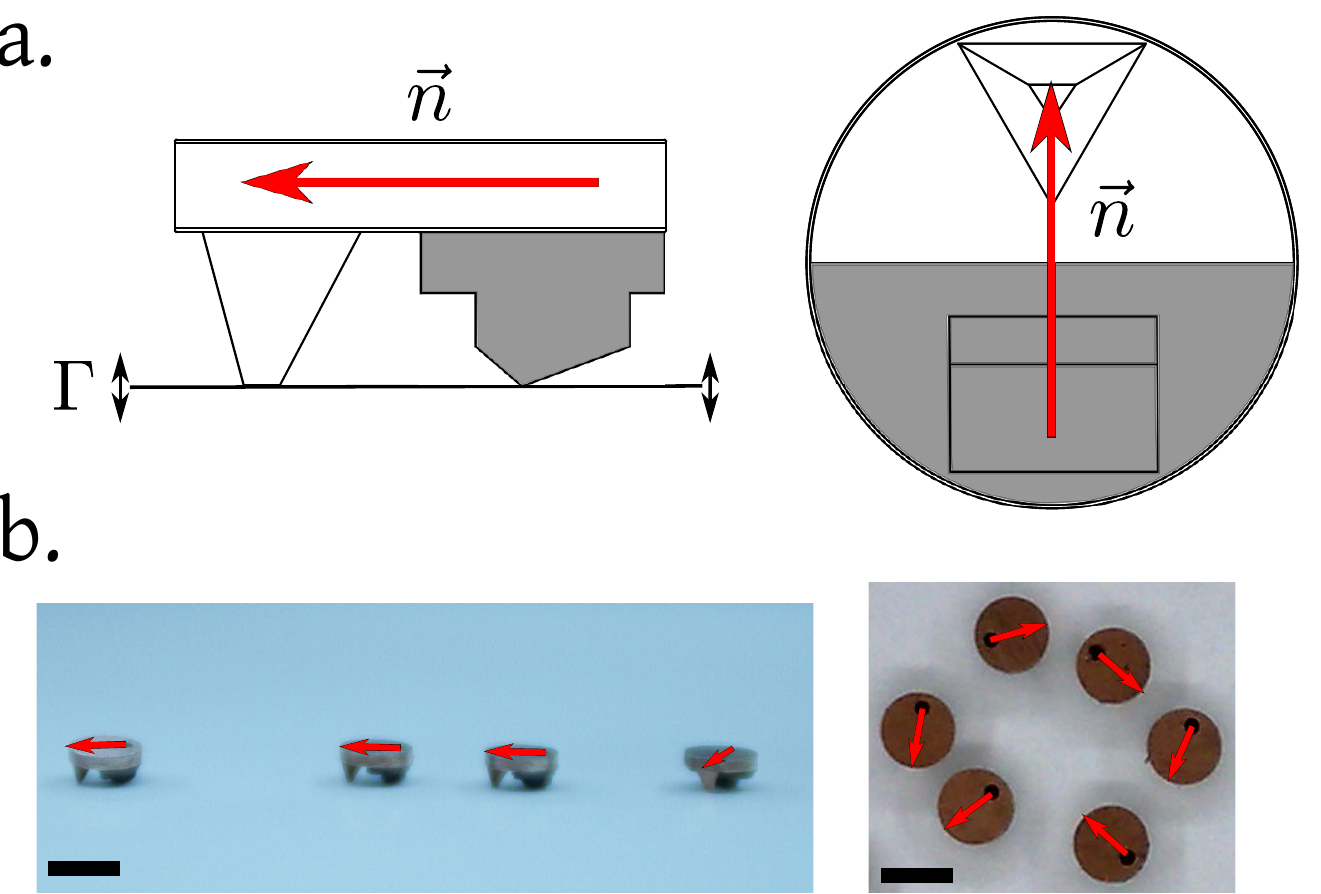}
\caption{ (color online) Self-propelled polar disks. 
(a) Side and bottom views of a polar disk with the built-in polarity $\vec{n}$. 
The white part of the particle is made of copper-berylium, while the grey part is made of nitrile.
(b) Side and top views of the polar disks with their respective polarities. The black scale bar is 4\,mm.
}
\label{fig:spp}
\end{figure}

To the best of our knowledge, all previous experiments looking at the
coherent displacement of objects colliding with a vertically-vibrated
horizontally-homogeneous substrate have used elongated particles 
\cite{yamada2003coherent,kudrolli2008swarming}. In other words,
the axis necessary to break rotational symmetry was simply encoded in 
the shape. In such a case, collisions directly induce alignment.
Here, by contrast, we use essentially circular particles: they consist of a 
top metallic disk with two asymmetric ``legs''  aligned to give the object
a polar axis (Fig.~\ref{fig:spp}a). These particles interact strictly 
via their top circular hard parts, much like more conventional hard disks.
Their polarity lies in the contacts with the vibrating plate.

In detail, we micro-machined copper-beryllium disks of diameter 4\,mm and height 2 \,mm 
with an off-center leg to which a nitrile skate of hardness 90 Sh was glued at a 
diametrically opposite position (Fig.~\ref{fig:spp}a). 
The skate is overmoulded to the copper-beryllium thanks to two polymeric layers, 
one adhesive to the metal (THIXON(TM) 520 Adhesive, Rhom and Haas), 
the other to the nitrile (THIXON P-11/0285SPL/25KG, Rhom and Haas). 
These two ``legs'', which have different mechanical response under vibration, 
endow the particles with a polar axis which can be determined 
from above thanks to a black spot located on their top (Fig.~\ref{fig:spp}b). 

In the following we also use plain disks (same metal,
diameter and height) as a control to ensure that all the effects reported 
here are indeed due to the ``self-propulsion'' induced by the built-in polarity.

\subsection{Data acquisition}

The position and polarity of the disks were recorded by a CCD camera 
with a spatial resolution of 1728x1728 pixels. 
The acquisition rate was set to 20\,Hz or 25\,Hz, i.e. about five times slower
than the shaking frequency of 115\,Hz. 
The ratio of the vibration frequency and the data acquisition rate sets the 
value of $\tau_0$, the smallest time increment accessible.
The trajectory of each particle was then reconstructed using 
standard tracking software. 
In the following, the time unit is set to be the vibration period 
and the unit length is the particle diameter. Within these units, 
the resolution on the position $\vec{r}$ of the particles is better than 0.05, 
and that on the orientation $\vec{n}$ is of the order of 0.05 rad.

\section{Single particle motion}
\label{single}

Our polar disks, once vibrated at $f = 115$\,Hz in our apparatus,
undergo fairly coherent, persistent
motion of the particles, over some range of easily accessible amplitudes $\Gamma=A_0 \omega^2/g$ with $\omega=2\pi f$.

\subsection{Typical behavior}

\begin{figure}[t!]
\centering
\vbox{
\hbox{
\hspace{0.5mm}
\includegraphics[trim = 0mm 0mm 0mm 0mm,clip,width=3.8cm,height=3.8cm]{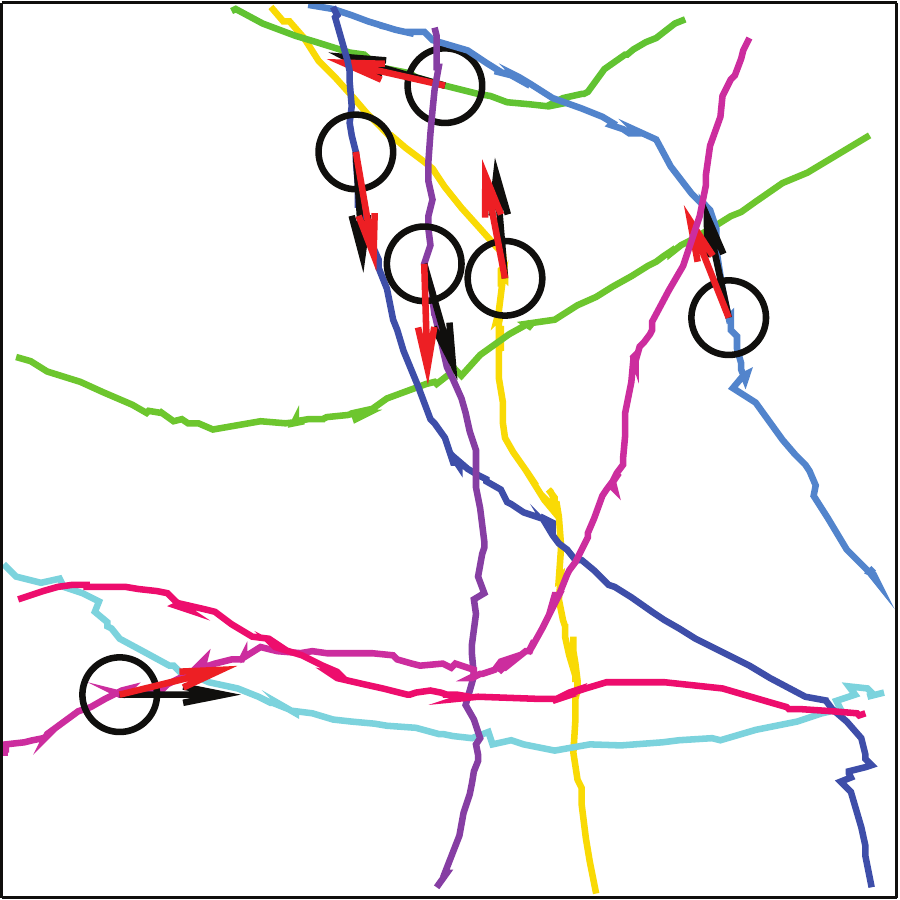}
\hspace{6mm}
\includegraphics[trim = 0mm 0mm 0mm 0mm,clip,width=3.8cm,height=3.8cm]{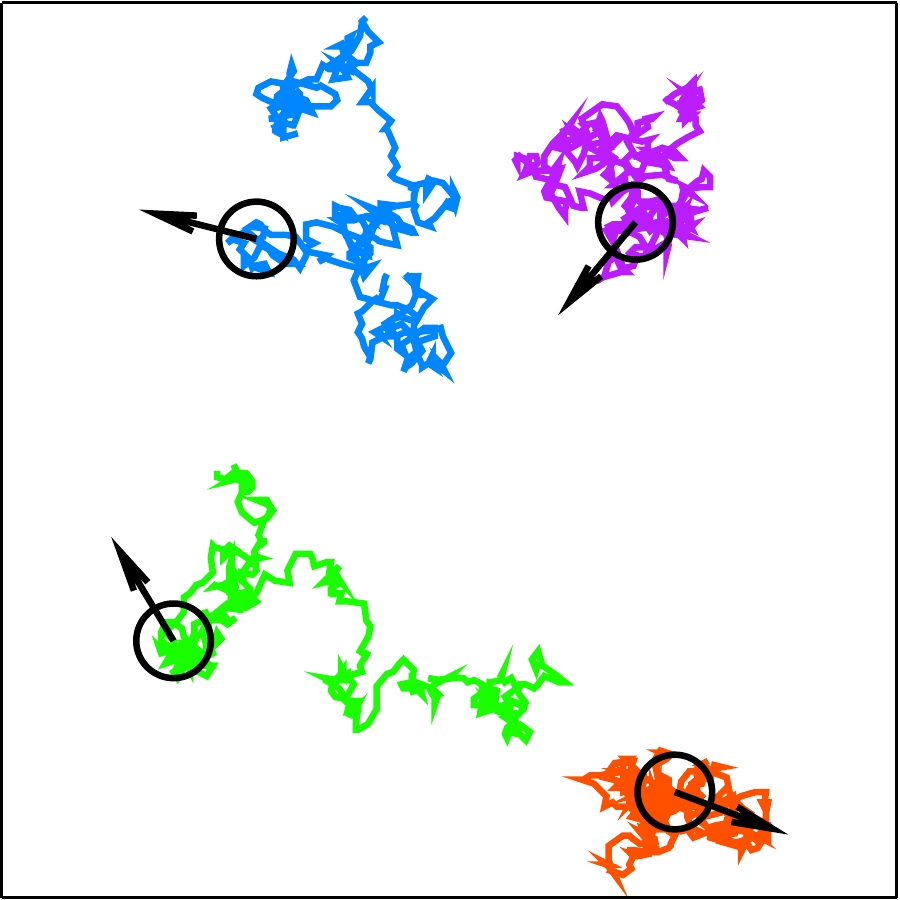}}
\hbox{\hspace{2cm}(a)\hspace{4.1cm} (b)}
}
\vspace{2mm}
\vbox{
\hbox{
\includegraphics[clip,width=4.25cm,height=4.2cm]{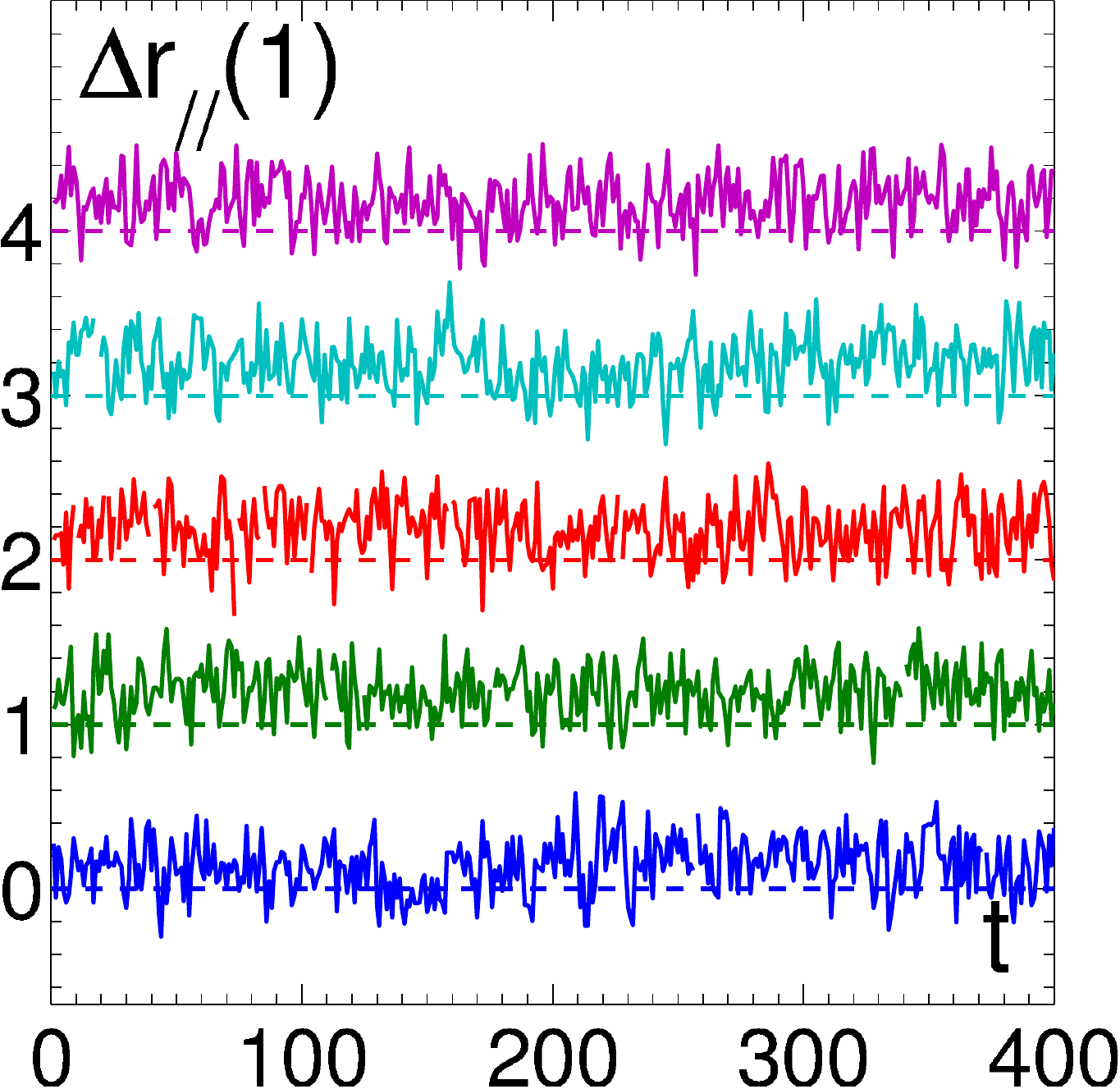}
\includegraphics[clip,width=4.25cm,height=4.2cm]{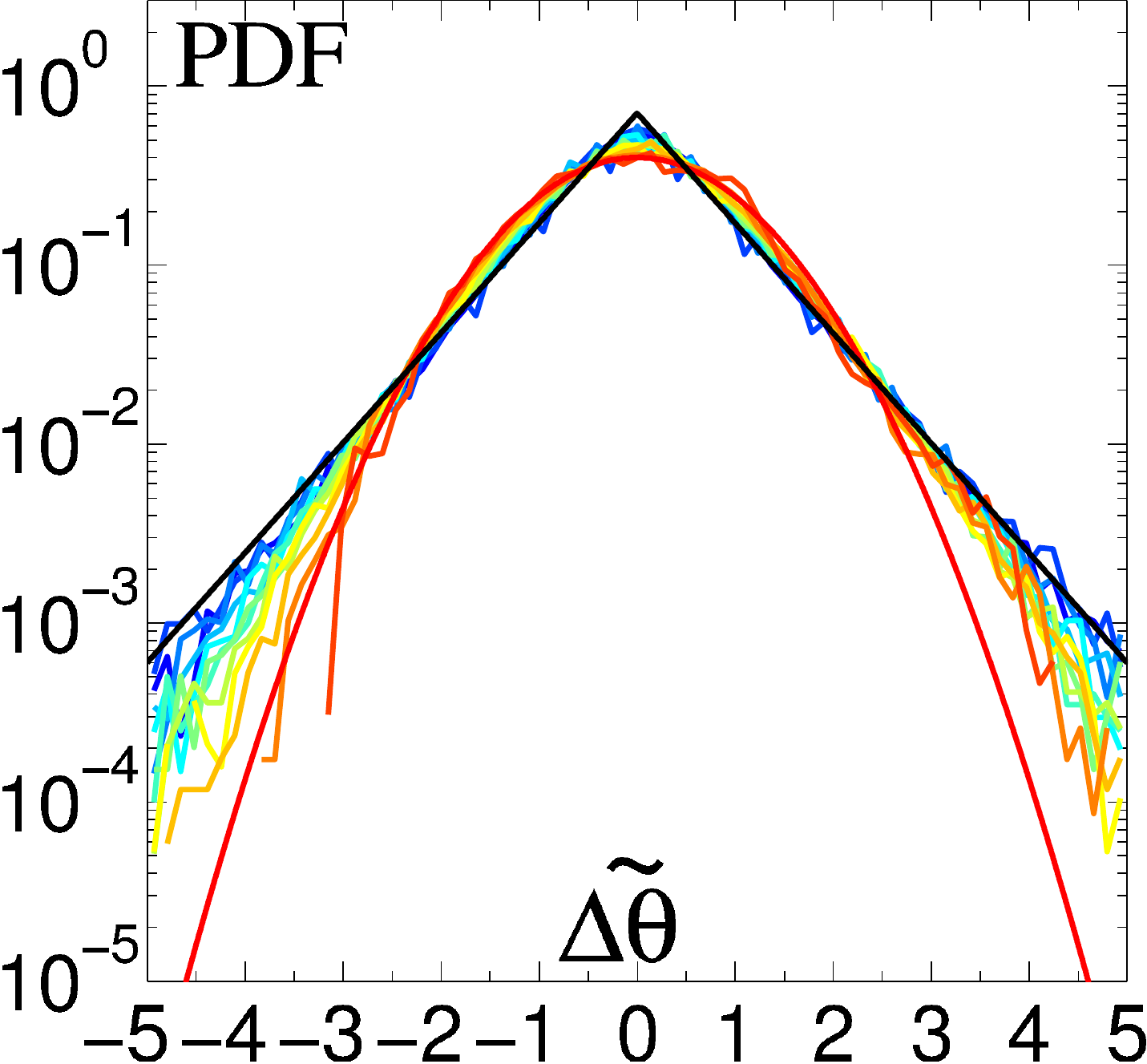}}
\hbox{\hspace{2cm}(c)\hspace{4.1cm} (d)}
}
\vspace{2mm}
\vbox{
\hbox{
\hspace{-5mm}
\includegraphics[clip,width=4.5cm, height=4.2cm]{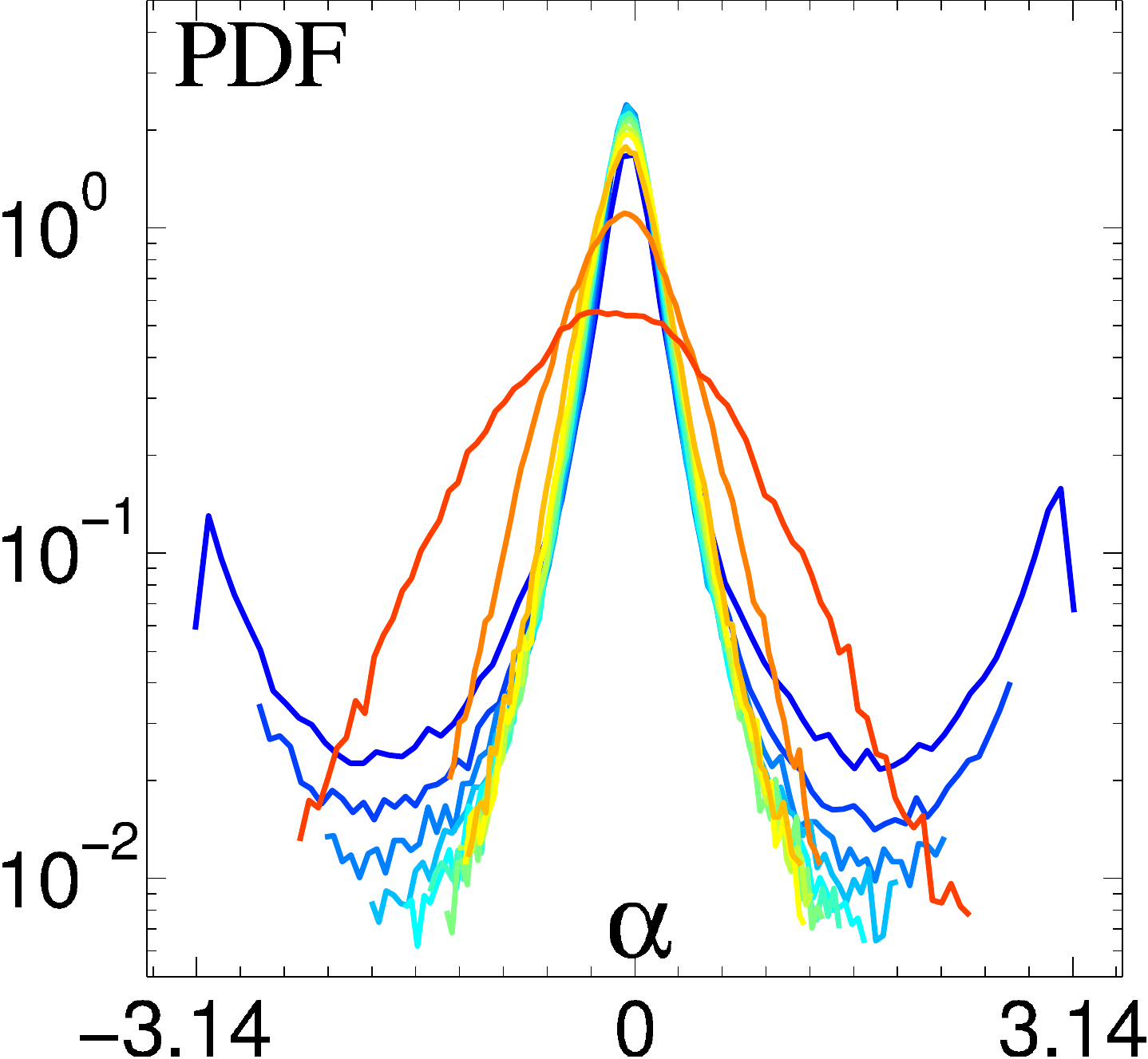}
\hspace{1mm}
\includegraphics[clip,width=4.1cm, height=4.2cm]{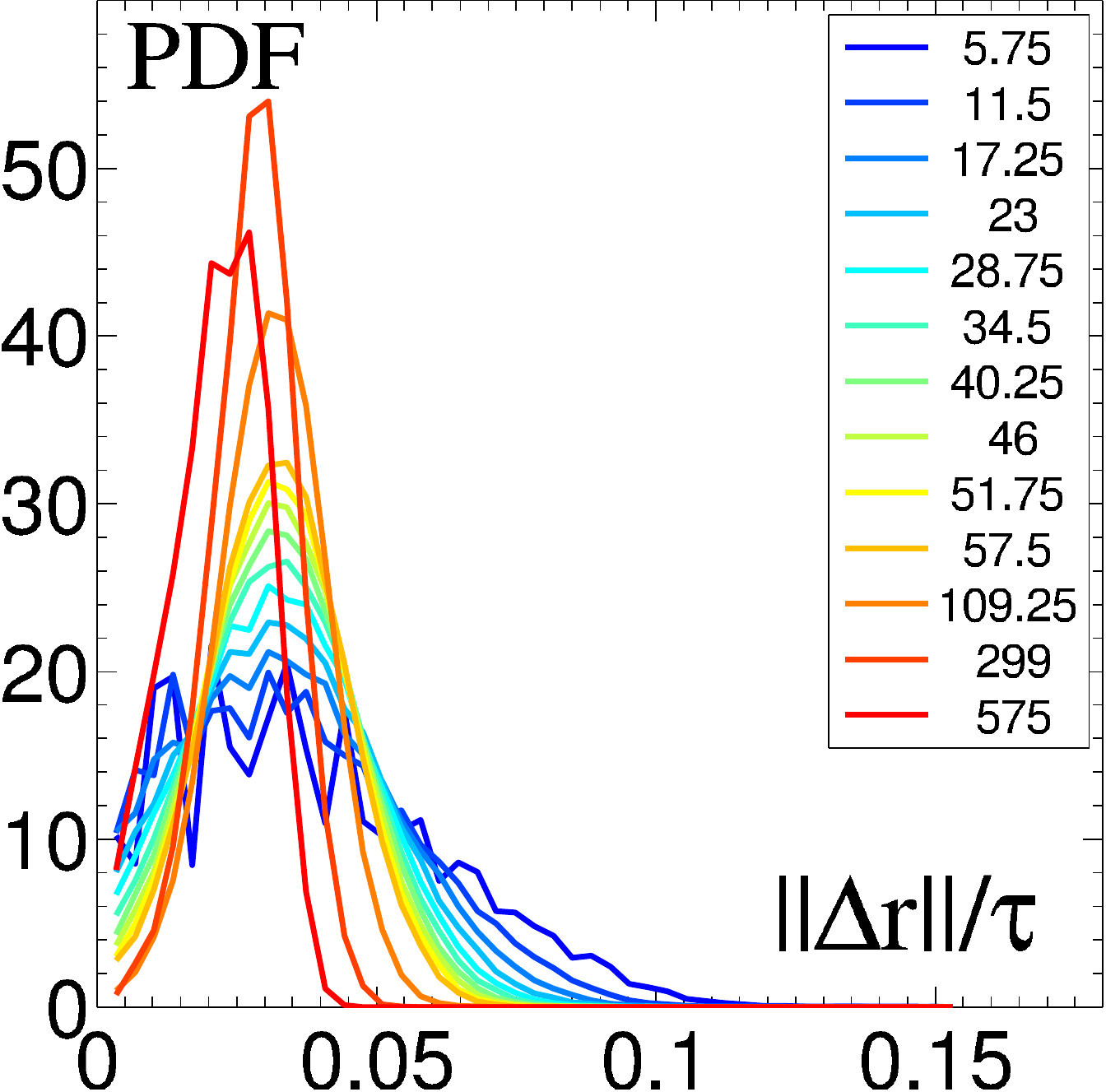}}
\hbox{\hspace{2cm}(e)\hspace{4.1cm} (f)}
}
\caption{ (color online) Statistical properties of the individual motion for $\Gamma=2.8$.
(a) Sample trajectories of the self-propelled particles. The red arrows indicate the instantaneous orientation of the polarity. The black arrows indicate the orientation of the displacement between two successive frames.
(b) Sample trajectories of the isotropic particles. The black arrows indicate the orientation of the displacement between two successive frames.
(c) Five time series of the displacement component along the polarity. The signals have been shifted for clarity. The dotted lines indicate the zero. One clearly observes negative events, corresponding to backward motion.
(d) Distribution of the reduced increments of the polarity orientation $\widetilde{\Delta\theta}(\tau)$ over a time $\tau$ for increasing $\tau$.
(e) Distribution of the angle $\alpha(\tau)$ between the displacement vector $\Delta \vec{r}(t,\tau)$ and the polarity $\vec{n}(t)$ for increasing $\tau$.
(f) Distribution of $|\Delta\vec{r}(t,\tau)|/\tau$, the displacement over a time $\tau$, normalized by $\tau$ for increasing $\tau$.
The color code in frames (d,e,f) is indicated by the legend in frame (f).
}
\label{fig:spp-stat}
\end{figure}

The single particle motion, which  arises from the interplay of collision/friction 
of the two legs with the bottom plate, and of the metallic disk with the top plate,
is too complicated to be described quantitatively. In the following, we only
discuss its projection on the horizontal plane, ``forgetting'' about the 
three-dimensional nature of the events at its origin.
Typical two-dimensional trajectories of isolated particles
appear smooth and rather straight (Fig.\ref{fig:spp-stat}a), 
as opposed to those of the control plain disks (Fig.\ref{fig:spp-stat}b).
A closer inspection, though, reveals that our polar particles typically
move steadily forward for some time, but also intermittently stop or even
move backward briefly  (Fig.\ref{fig:spp-stat}c) and Supplementary movie \cite{epaps1}).

The directed, persistent motion of particles is recorded in the (discrete-)time series 
of their position $\vec{r}(t)$ and polarity  $\theta(t)$.
The polarity is very persistent in time: the distribution of 
$\Delta\theta(\tau)=\theta(t+\tau)-\theta(t)$, its increment over time $\tau$,
is essentially exponential and sharply peaked at zero for $\tau=\tau_0$. 
Moreover, the distribution of the {\it reduced} increment 
$\widetilde{\Delta\theta}(\tau)=\Delta\theta(\tau)/ \langle\Delta\theta(\tau)^2\rangle^{1/2}$
remains roughly independent of $\tau$ up to fairly large values of $\tau$,
after which it gradually recovers the expected asymptotic Gaussian character
(Fig.~\ref{fig:spp-stat}d). The crossover value of $\tau$ (of the order of 100) 
can be taken as a rough estimate of the persistence time of polarity.

By and large, the displacement of the particles is overwhelmingly along their 
polarity axis: the distribution function of the angle $\alpha(\tau)$ between 
$\Delta\vec{r}(t,\tau)=\vec{r}(t+\tau)-\vec{r}(t)$, 
the displacement vector over time $\tau$, and $\theta(t)$ is peaked around zero.
Strikingly, this is all the more so as $\tau$ is large:
The backward motion events apparent in the 
secondary peaks at $\pm\pi$ present at small $\tau$ average out 
as $\tau$ is taken larger, an indication that during most of these ``backward events'' 
the polarity remains largely unchanged (Fig.~\ref{fig:spp-stat}e). 
Again, this is only true up to some crossover value of $\tau\simeq 100$
beyond which the distribution of $\alpha(\tau)$ must gradually become flat.
For $\tau$ larger than the crossover, one indeed observes a widening of the distribution.
This displacement mostly along the polarity axis is performed at a 
fairly well-defined speed: for not too large $\tau$, 
the distribution of $|\Delta\vec{r}(t,\tau)|/\tau$ is peaked around a most probable value. 
Increasing $\tau$ from $\tau_0$ to 
about the crossover time mentioned above, the distribution 
keeps the same most probable value and gets narrower and narrower
(Fig.~\ref{fig:spp-stat}f). The well-defined most probable value is thus nothing but the
average speed $\langle v \rangle$.  
This indicates again that over these timescales our particles essentially go straight.
The distribution is then essentially Gaussian. For timescales larger than the crossover, 
there is  of course a shift of the ``speed'' towards lower values as expected 
when the particles enter the long-time, uncorrelated, diffusive regime.

\subsection{Influence of the vibration amplitude $\Gamma$}

\begin{figure}[t!]
\centering
\vbox{
\hbox{
\includegraphics[clip,width=4.1cm,height=4.1cm]{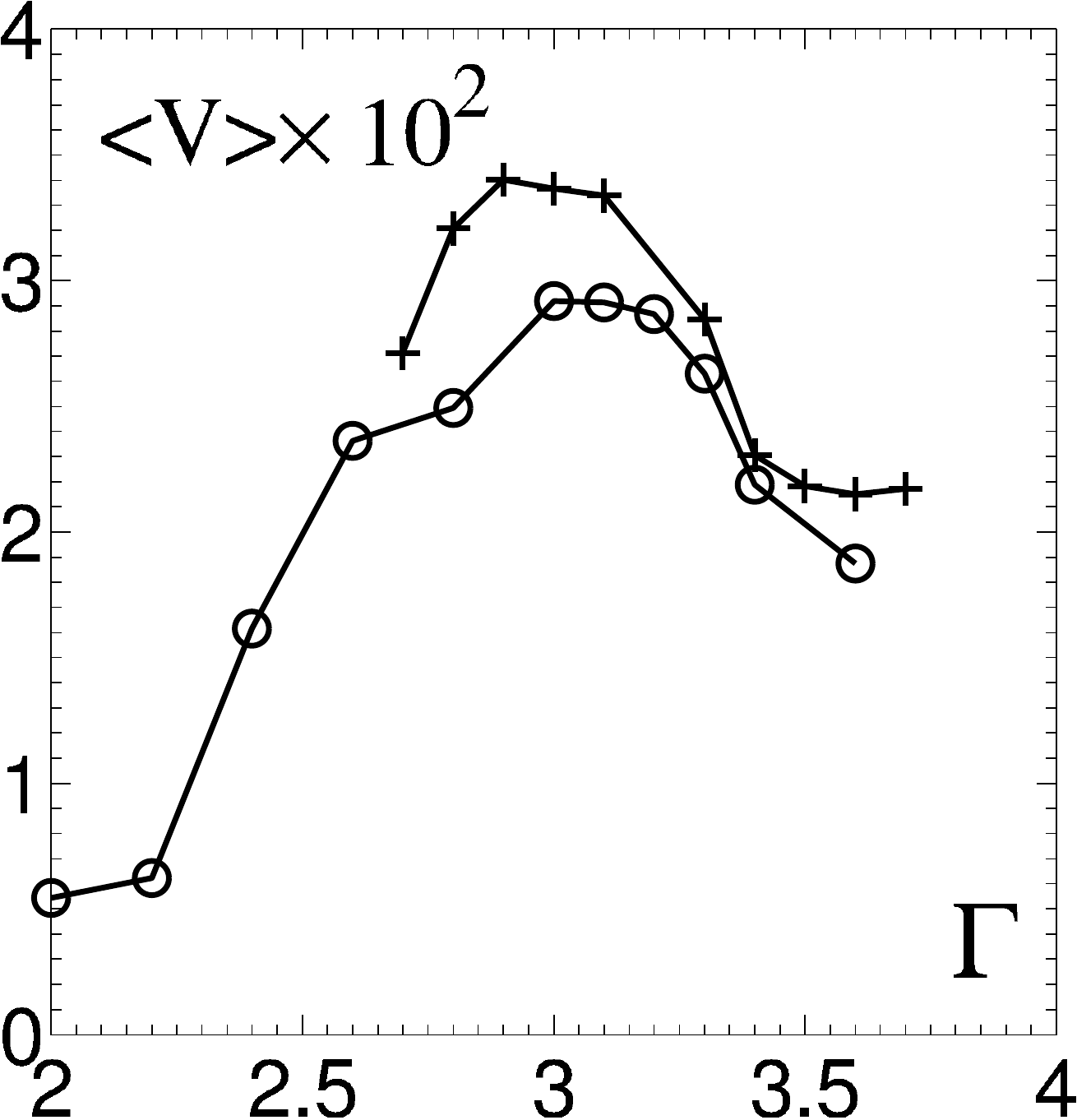}
\includegraphics[clip,width=4.2cm,height=4.0cm]{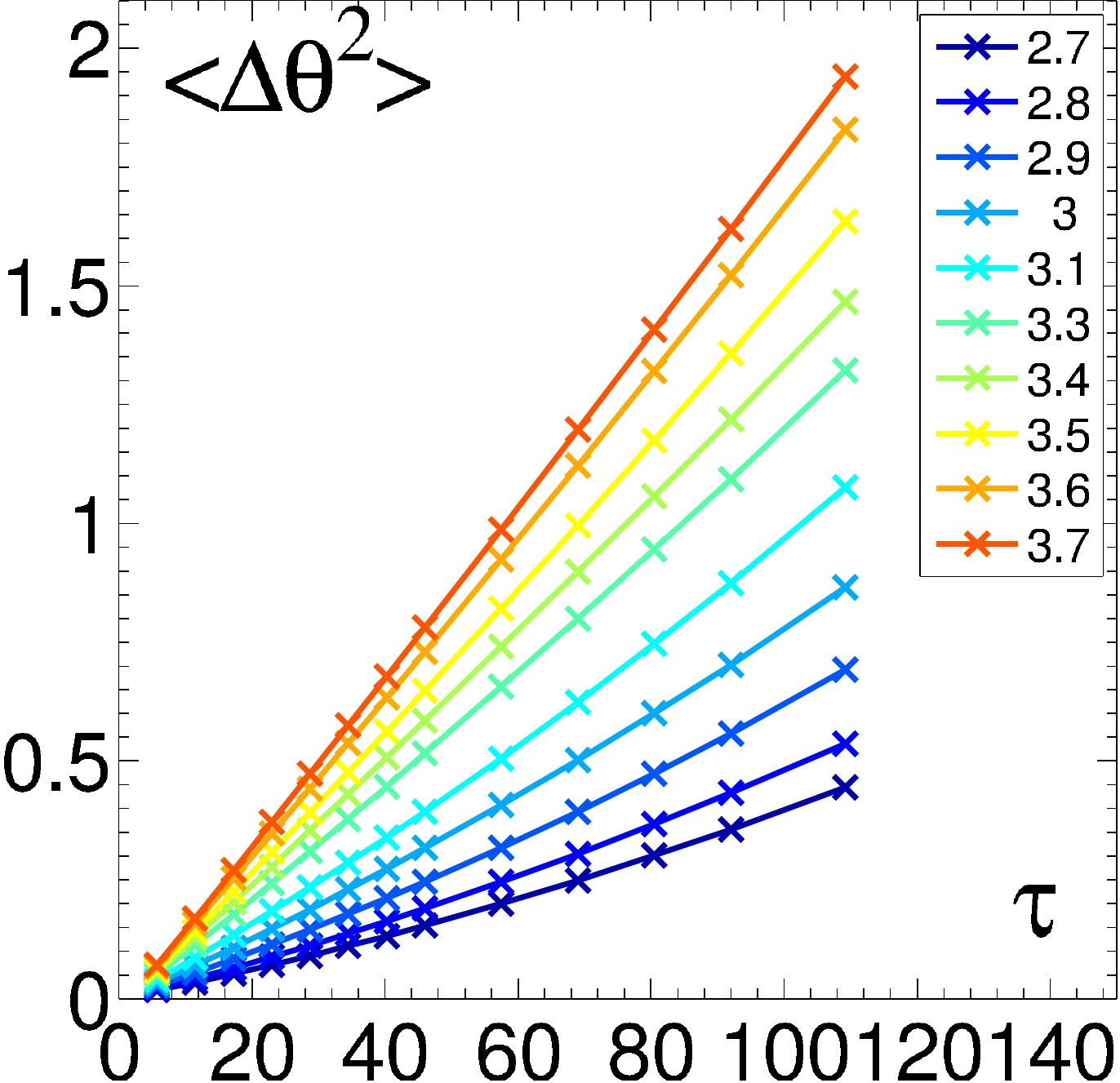}}
\hbox{\hspace{2cm}(a)\hspace{3.9cm} (b)}
}
\vbox{
\hbox{
\hspace{-4mm}
\includegraphics[clip,width=4.4cm,height=4.2cm]{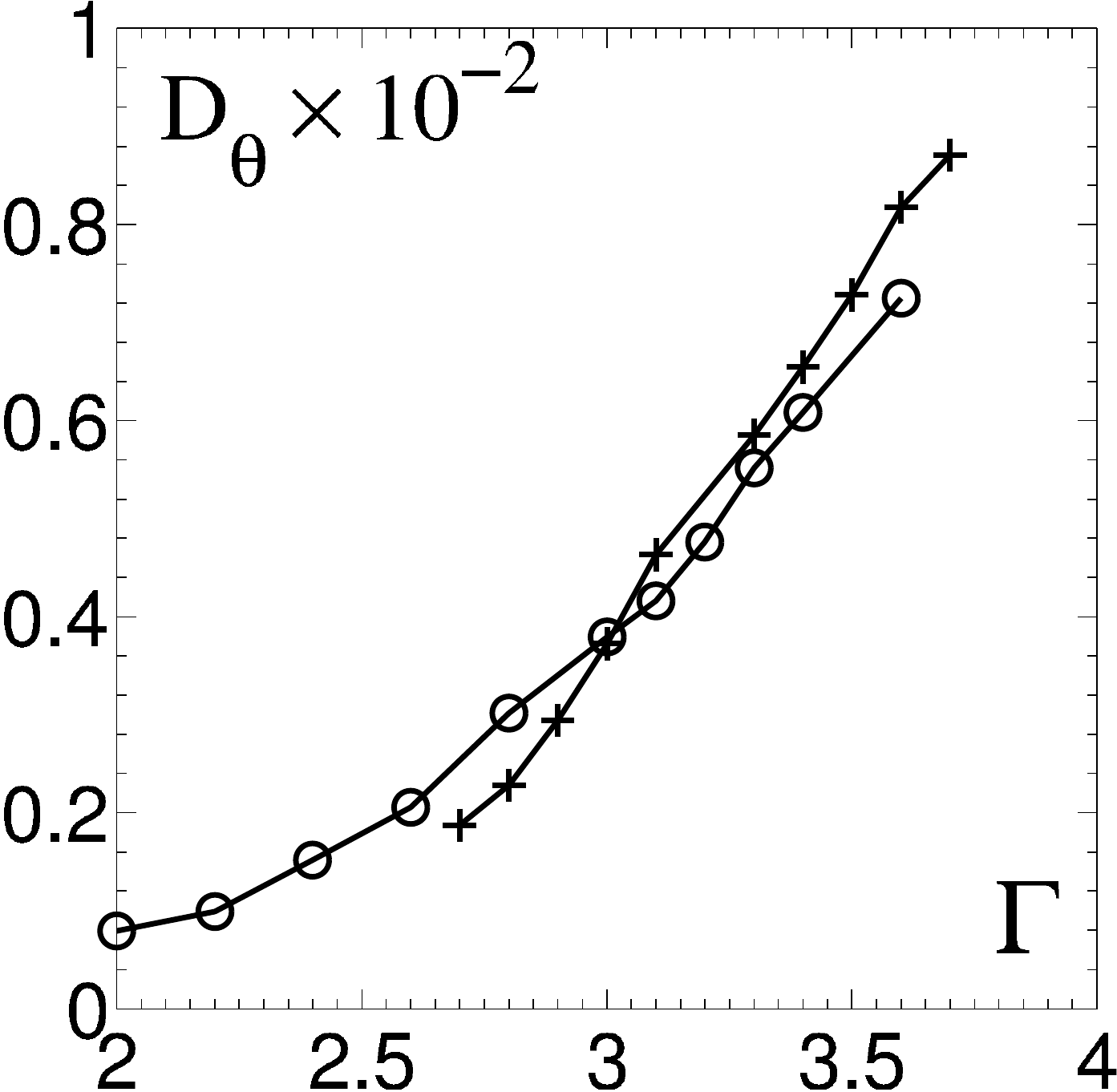}
\includegraphics[clip,width=4.2cm,height=4.1cm]{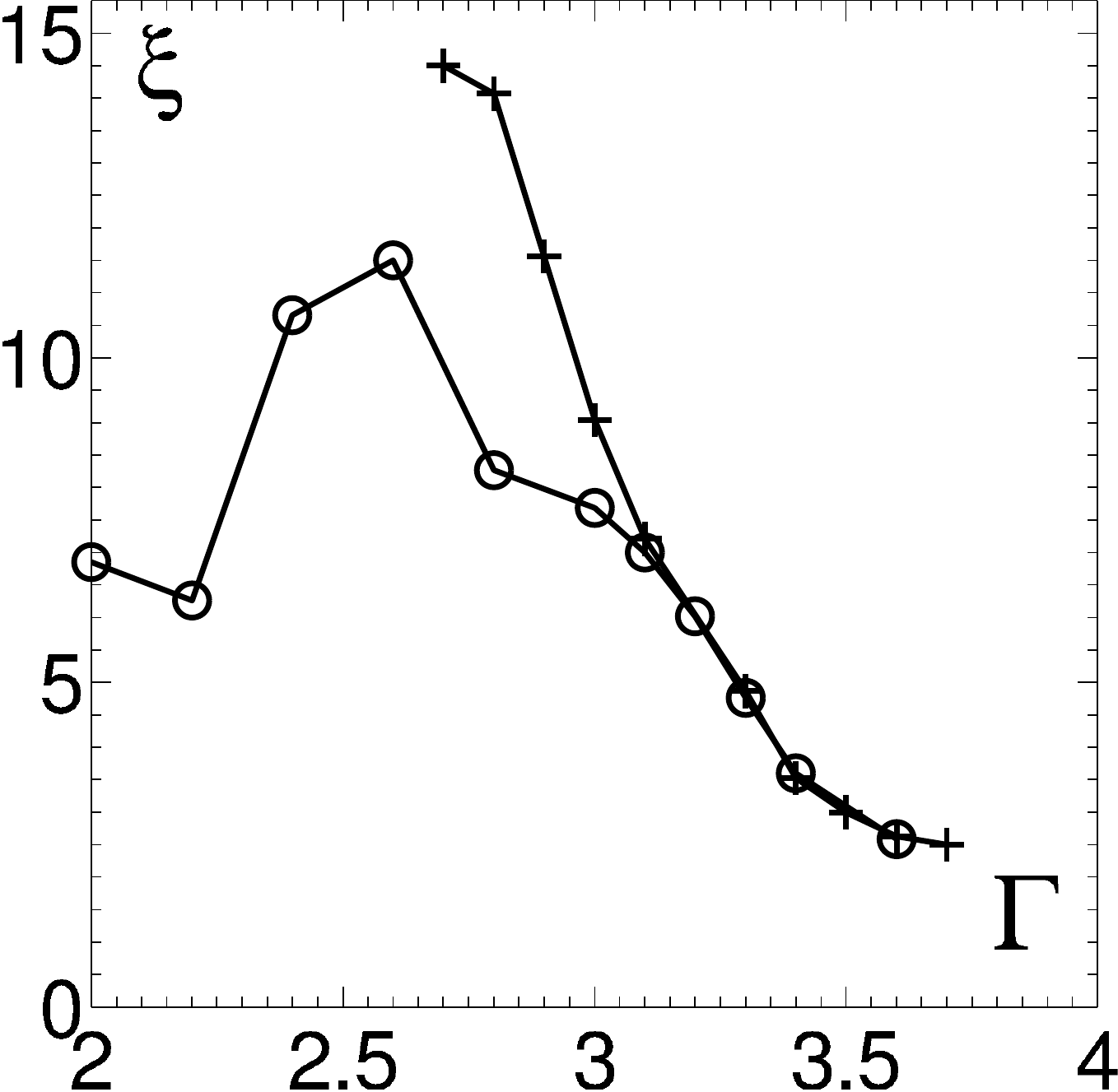}}
\hbox{\hspace{2cm}(c)\hspace{3.9cm} (d)}
}
\caption{(color online) Self propulsion 
(a) Characteristic speed $\langle v\rangle$ defined by the most probable value of $\| \Delta\vec{r}(\tau)/\tau$ vs. $\Gamma$ for $\tau\le100$. 
(b) Mean square angular increment computed on a lag time $\tau$ versus $\tau$, for vibration amplitude $\Gamma \in [2.7-3.7]$. 
(c) Diffusion coefficient $D_{\theta}$ of the orientation of the polarity vs. $\Gamma$.
(d) Persistence length $\xi=\langle v\rangle/D_{\theta}$ vs. $\Gamma$.
Symbols $(+)$ and $(\circ)$ indicate two different experimental runs with different experimental conditions, such as ambient humidity, fine positioning of the top glass plate, etc.... The data analyzed in this paper for the small system correspond to the run labelled $(+)$.
}
\label{fig:param}
\end{figure}

We have seen above that our polar disks can be faithfully described, over scales which average out
the stopping and backward events, as moving at a well-defined finite speed
 $\langle v \rangle$ while being subjected to weak rotational diffusion.
A direct and accurate measure of the (rather long) persistence length/time of the 
trajectories of our particles via, say, the time decay of the 
autocorrelation of their polarity, 
is rendered difficult by the relatively small size of our system. The 
data presented in Fig.~\ref{fig:spp-stat} were obtained in a dish of diameter 40
vibrated at an amplitude $\Gamma=2.8$. In such conditions,
the rather straight trajectories will hit the wall long before their have turned 
enough to yield a significant decay of the polarity autocorrelation. 
To overcome this difficulty, we used the distribution of (normalized)
polarity increments as shown in Fig.~\ref{fig:spp-stat}d. Being independent of 
$\tau$ at small $\tau$, and the mean square angular increment being linear
in $\tau$ (Fig.~\ref{fig:param}a), it allows to define the rotational diffusion constant $D_\theta$
as its (half-)variance, and thus the persistence length as $\xi=\langle v\rangle / D_\theta$.

The influence of $\Gamma$ on  $V_m$,  $D_\theta$, and $\xi$ is described in Fig.~\ref{fig:param}. 
First of all, we observe a rather sudden drop in self-propulsion
when  $\Gamma$ is decreased (Fig.~\ref{fig:param}a):  for small amplitude, the drive is too weak and 
our polar disks do not move much. Over the range  $\Gamma\in [2.7, 3.8]$, on the other hand,
the average speed varies little. In contrast, the rotational diffusion constant $D_\theta$, which is easily
captured from the root mean square angular increment computed on a lag $\tau$ (Fig.~\ref{fig:param}b), steadily 
increases with $\Gamma$, with an almost linear behavior in the $\Gamma$-range over which 
steady propulsion occurs  (Fig.~\ref{fig:param}c). This indicates that the main effect of 
$\Gamma$ is on the strength of the angular noise.
As a consequence, the persistence length $\xi$ becomes large as $\Gamma$ is decreased 
within the steady propulsion range $[2.7, 3.8]$, but falls back for too small $\Gamma$ values 
(Fig.~\ref{fig:param}d).

\section{Binary collisions}
\label{binary}

\begin{figure*}[t!] 
\vbox{
\hbox{
\includegraphics[clip,width=4.2cm,height=4.2cm]{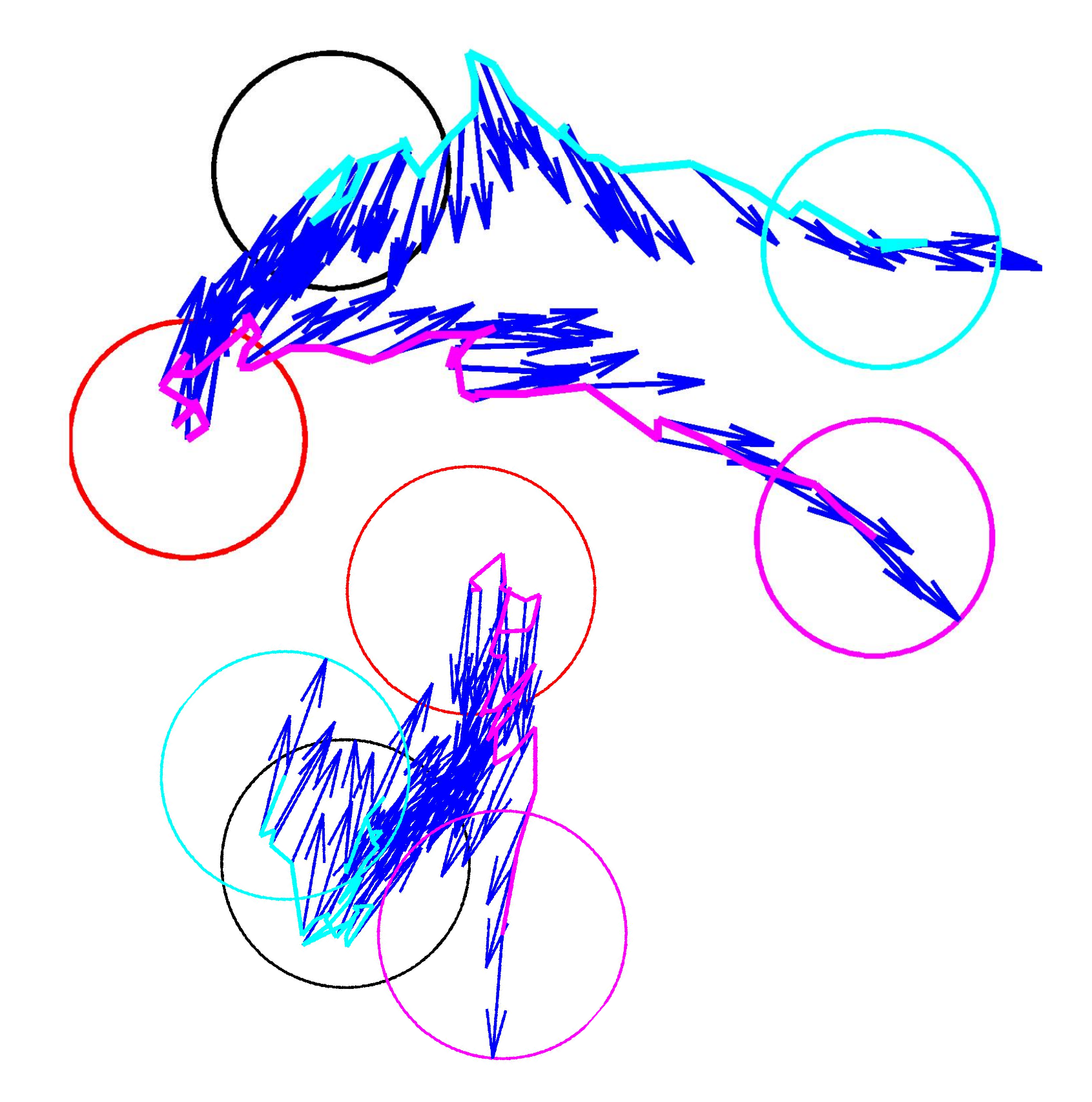}
\includegraphics[clip,width=9cm]{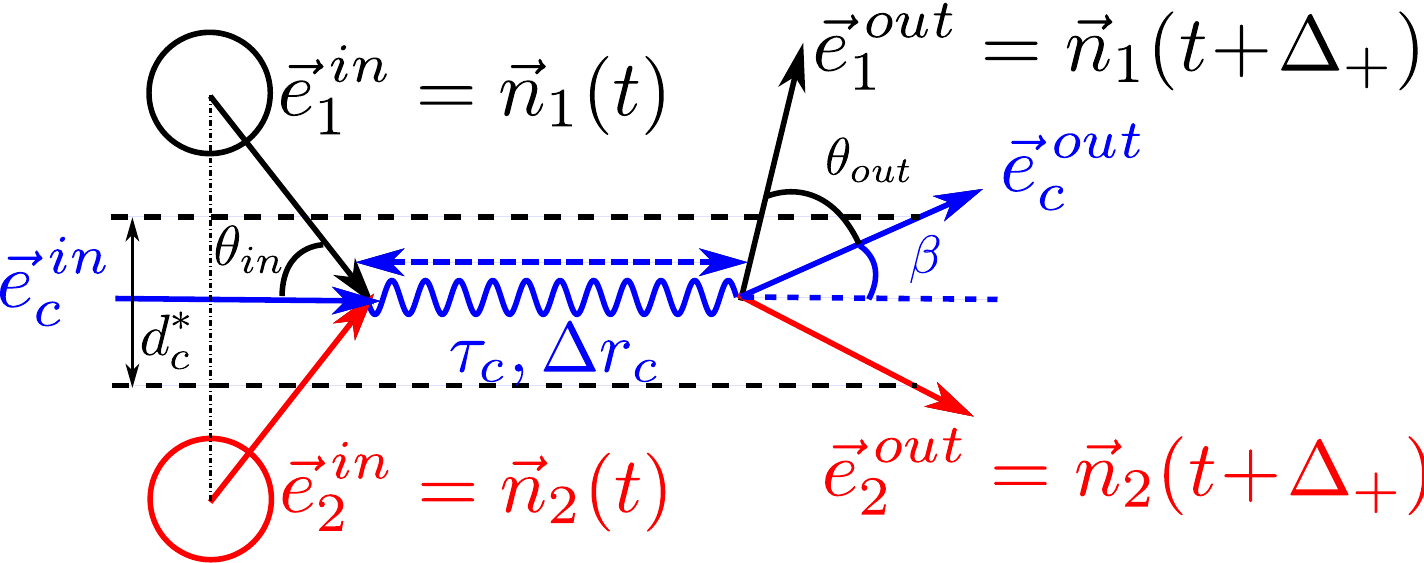}
\includegraphics[clip,width=4cm,height=4cm]{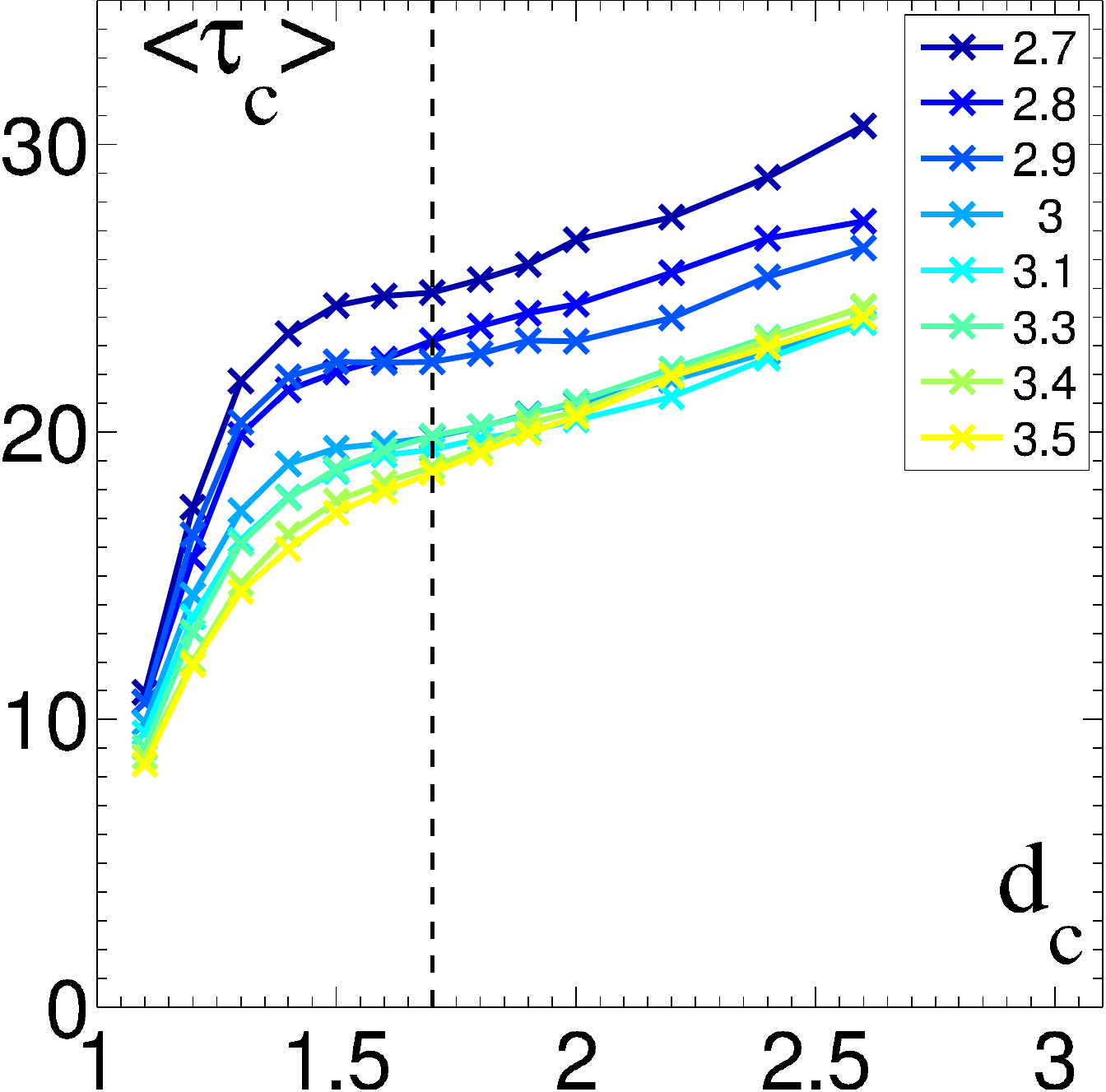}}
\hbox{\hspace{2cm}(a)\hspace{6cm} (b)\hspace{6.3cm} (c)}
}
\vspace{5mm}
\vbox{
\hbox{
\includegraphics[clip,width=5.4cm,height=5cm]{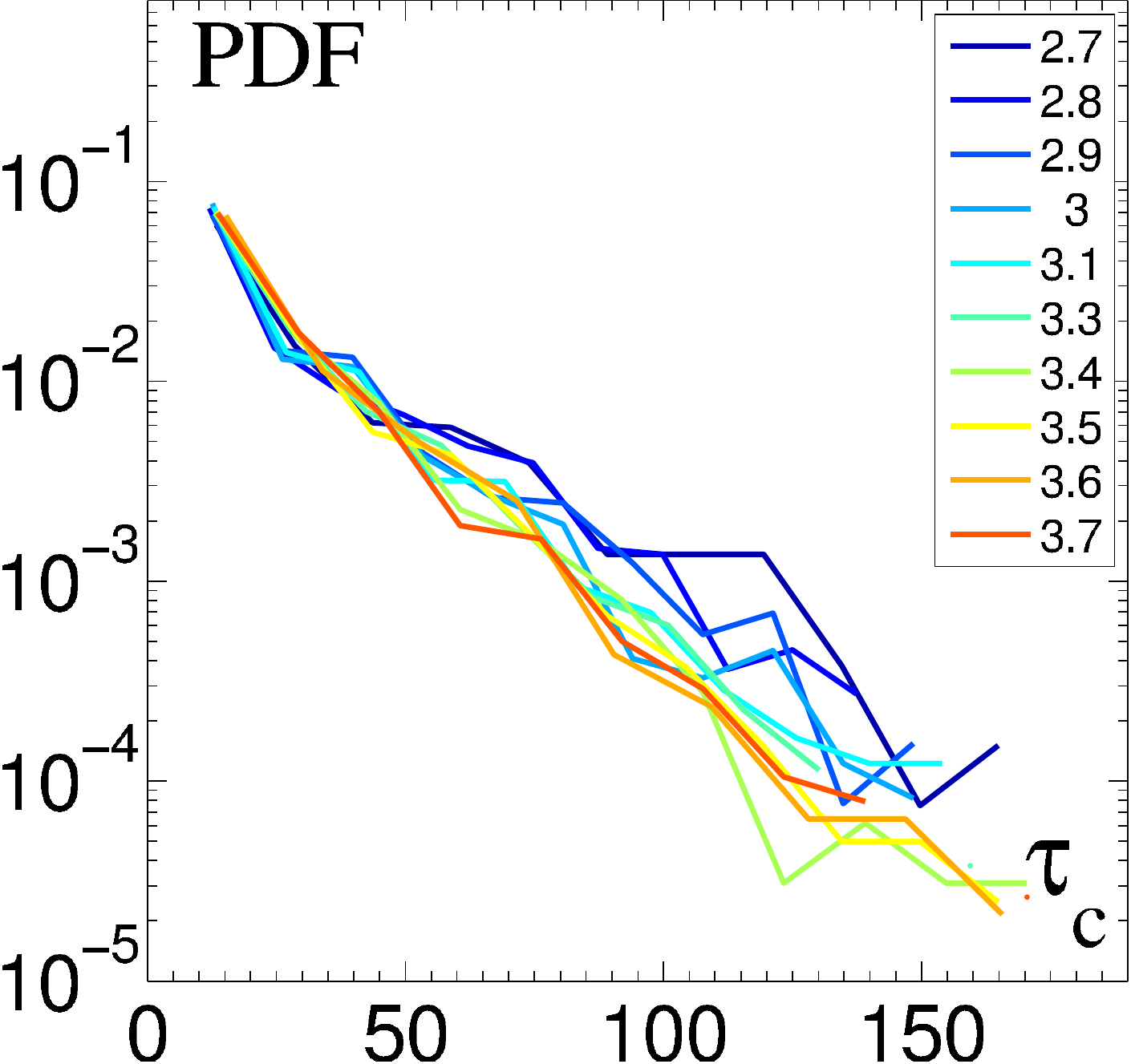}
\hspace{0.4cm}
\includegraphics[clip,width=5.4cm,height=5cm]{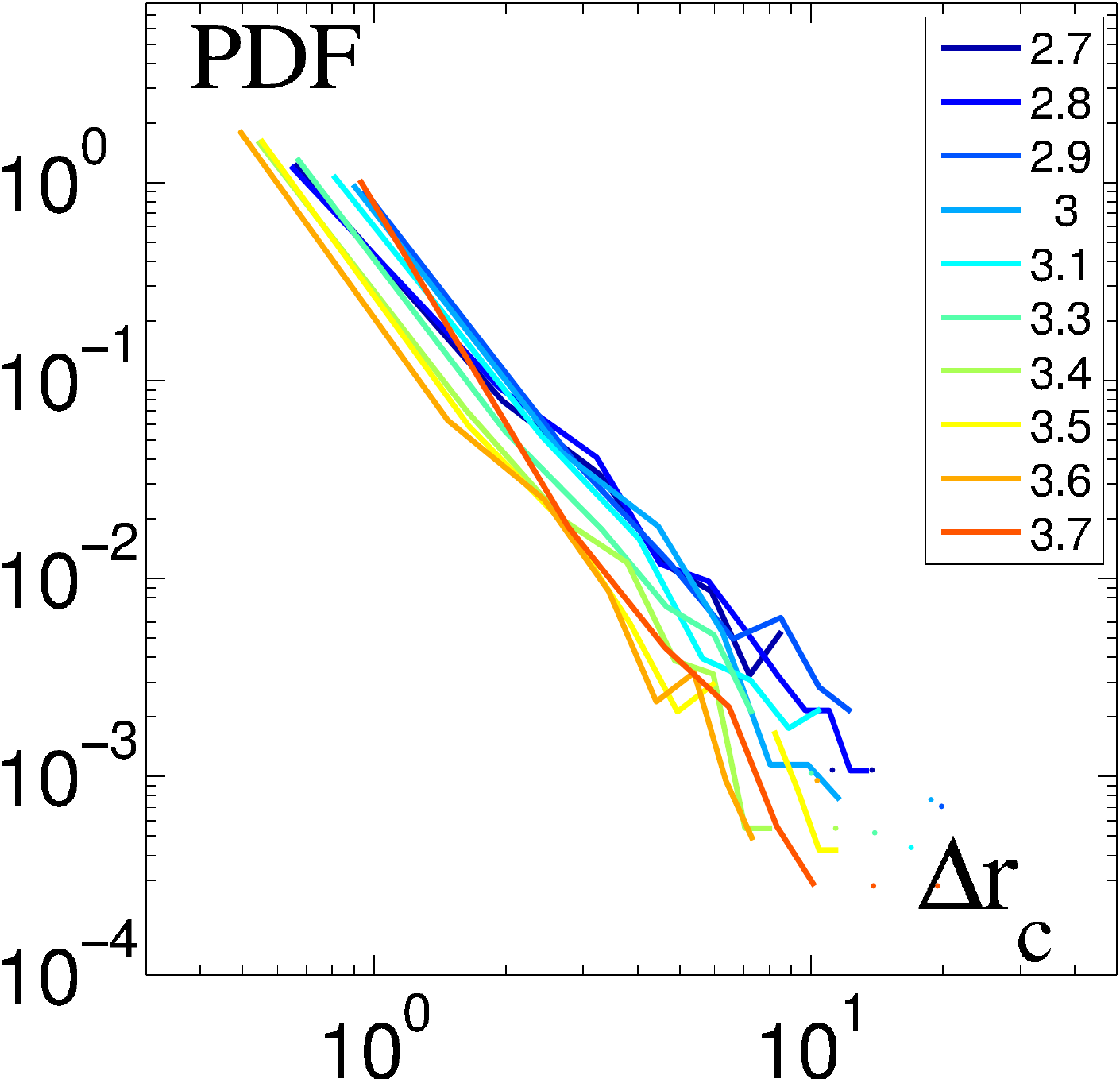}
\hspace{0.4cm}
\includegraphics[clip,width=5.4cm,height=5cm]{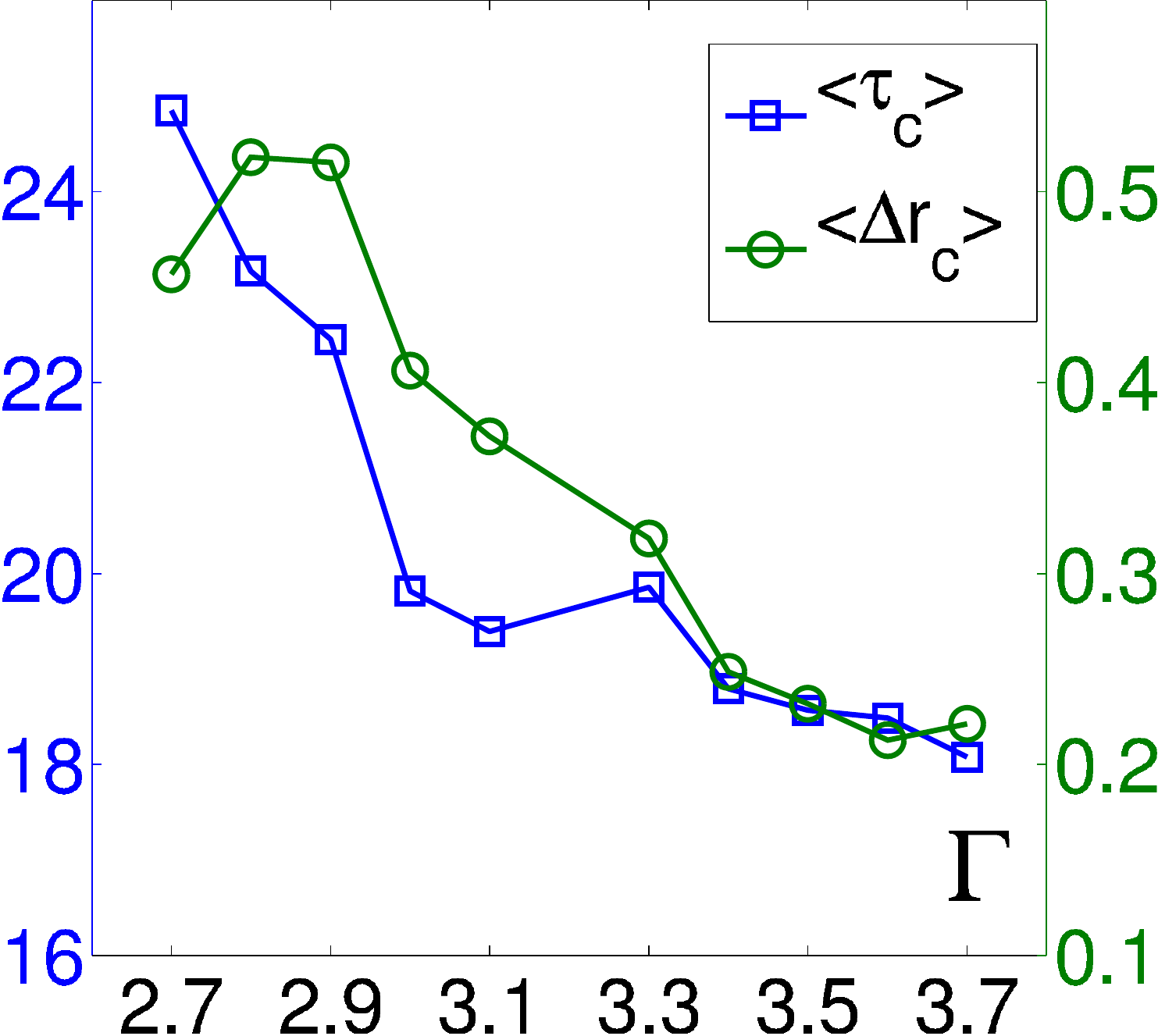}}
\hbox{\hspace{3cm}(d)\hspace{5cm} (e)\hspace{5.5cm} (f)}
}
\caption{(Color online) Collisions.
(a) Two samples of encounters exhibiting distinct effective alignment properties. In the case of the encounter on top, the two particles encounter  face to face, collide successively, while their polarities align, until the encounter ends at a relatively large distance from the first collision position.  In the case of the bottom one, the two particles again encounter face to face. However the particles polarities essentially do not change during the successive collisions and he encounter ends when the impact factor of the collision reduces to zero. At the end of the encounter the particles escape back to back.
(b) Sketch of an encounter with the definition of its geometrical properties.
(c) Mean duration of the encounters as a function of the distance $d_{\rm c}$ used as a criterion to define them, (color code as indicated in the legend); the vertical dash line indicate the criterion we have kept to define encounters.
(d) Distribution of encounter durations $\tau_c$, (color code as indicated in the legend).
(e) Distribution  of encounter lengths $\Delta r_{\rm c}$, (color code as indicated in the legend). 
(f) Average encounter duration $<\tau_c>, (\square)$ and length $<\Delta r_{\rm c}>, (o)$ as a function of $\Gamma$.}
\label{fig:coll-stat}
\end{figure*}

Having characterized the motion of our particles, we now turn to a statistical
description of their binary collisions using trajectories recorded when only few
particles are evolving in the system (typically 50 particles in a domain of
diameter $\sim40$). As already stated, most models for the collective motion
of objects moving on a substrate involve more or less explicit alignment rules.
For elongated, rod-like particles, inelastic collisions immediately lead to
alignment. Here, given that our particles interact by collisions of their
circular top part, no such direct alignment occurs. Rather, encounters
typically consist of multiple collisions each followed by some rebounds 
(Fig.~\ref{fig:coll-stat}a and Supplementary movie \cite{epaps2}). 
It is thus by no means clear a priori that even some effective, average 
alignment takes place during collisions of two of our particles.

\subsection{Defining collisions, their duration, and their spatial extent}

Given that encounters of two of our particles typically involve many actual
collisions, sometimes occuring more frequently than our sampling rate, we
studied the statistics of encounters defined by the total space-time event
during which two particles, initially far apart, have had their centers 
stay closer from each other than some threshold distance $d_{\rm c}$. To insure 
that these events are ``real'' collisions, this distance criterion was
supplemented by the condition that the polarities of the two particles
at the beginning of the encounter indicate that their distance is
likely to decrease.

Obviously, $d_{\rm c}$ must be chosen larger than 1, but not too large otherwise
encounters may not cease. To use a unique, objective value of  $d_{\rm c}$,
we calculated $\langle\tau_{\rm c}\rangle$, the mean duration of encounters,
as a function of  $d_{\rm c}$, over a set of a few thousands events. 
The resulting curve shows a plateau-like behavior in the range $d_{\rm c}\in [1.5-1.9]$,
the middle of which can be defined as being the inflection point $d^*_{\rm c}\simeq 1.7$ (Fig.~\ref{fig:coll-stat}c).
Furthermore, this well-defined value, around which the statistics of encounters
do not vary significantly, does not vary much when $\Gamma$ is 
changed within the useful range $[2.7,3.8]$. It is thus adopted in the following to define encounters 
(hereafter renamed collisions for simplicity) quantitatively.
Once collisions thus defined,  their spatial extension $\Delta r_{\rm c}$
can be estimated as the distance between the mid point of the particle centers at the
beginning and at the end of the collision (Fig.~\ref{fig:coll-stat}b)
We find the collision sizes distributed roughly algebraically with an
exponent of the order of 3, whereas the collision durations $\tau_{\rm c}$
are distributed roughly exponentially.
The influence of $\Gamma$ on these distributions remains rather weak,
with experimentally-observed means decreasing slowly with $\Gamma$
(Fig.~\ref{fig:coll-stat}d-f).

\begin{figure}[t!]
\centering 
\vbox{
\hbox{
\includegraphics[clip,width=4.1cm,height=4cm]{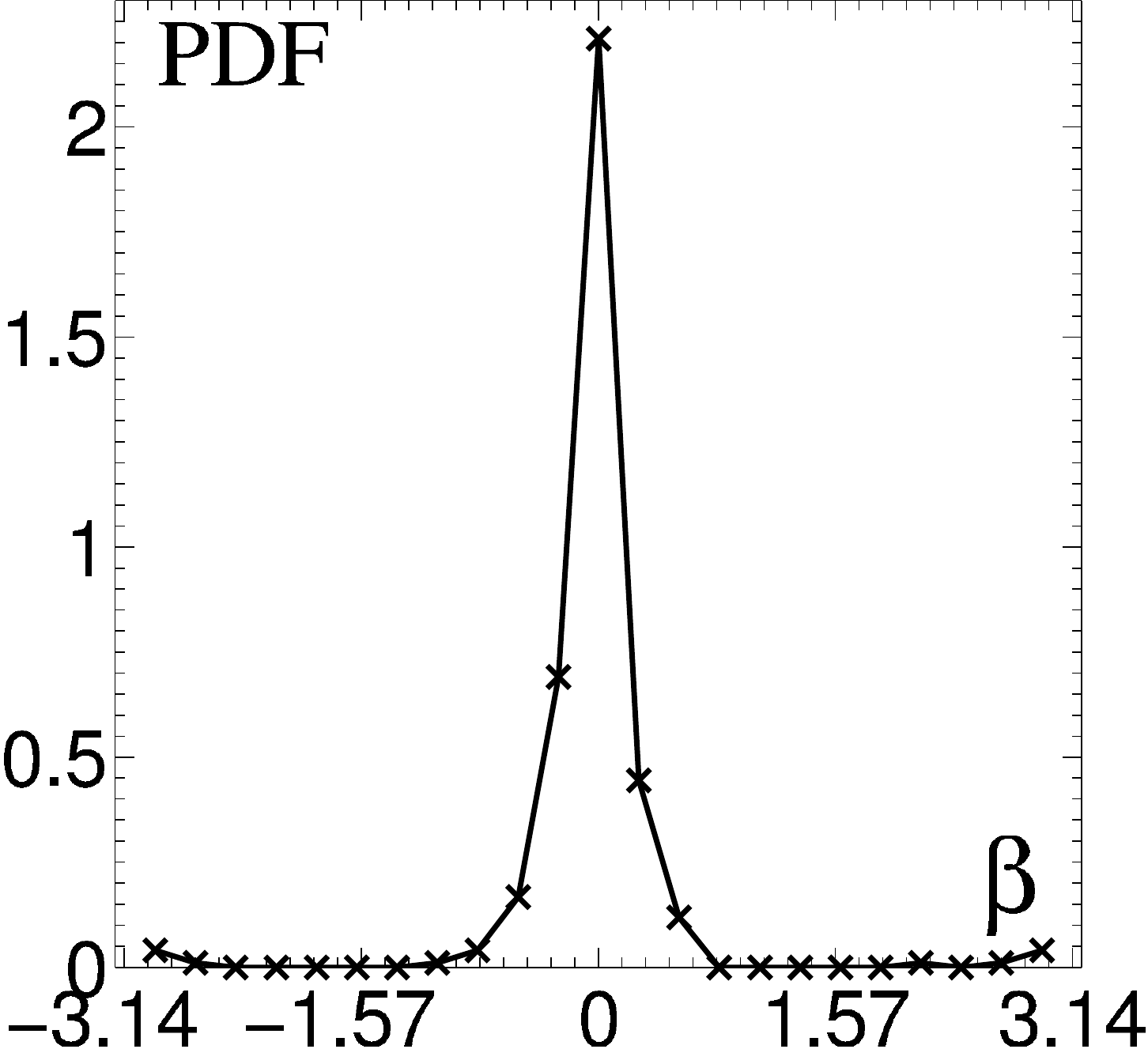}
\includegraphics[clip,width=4.4cm,height=4cm]{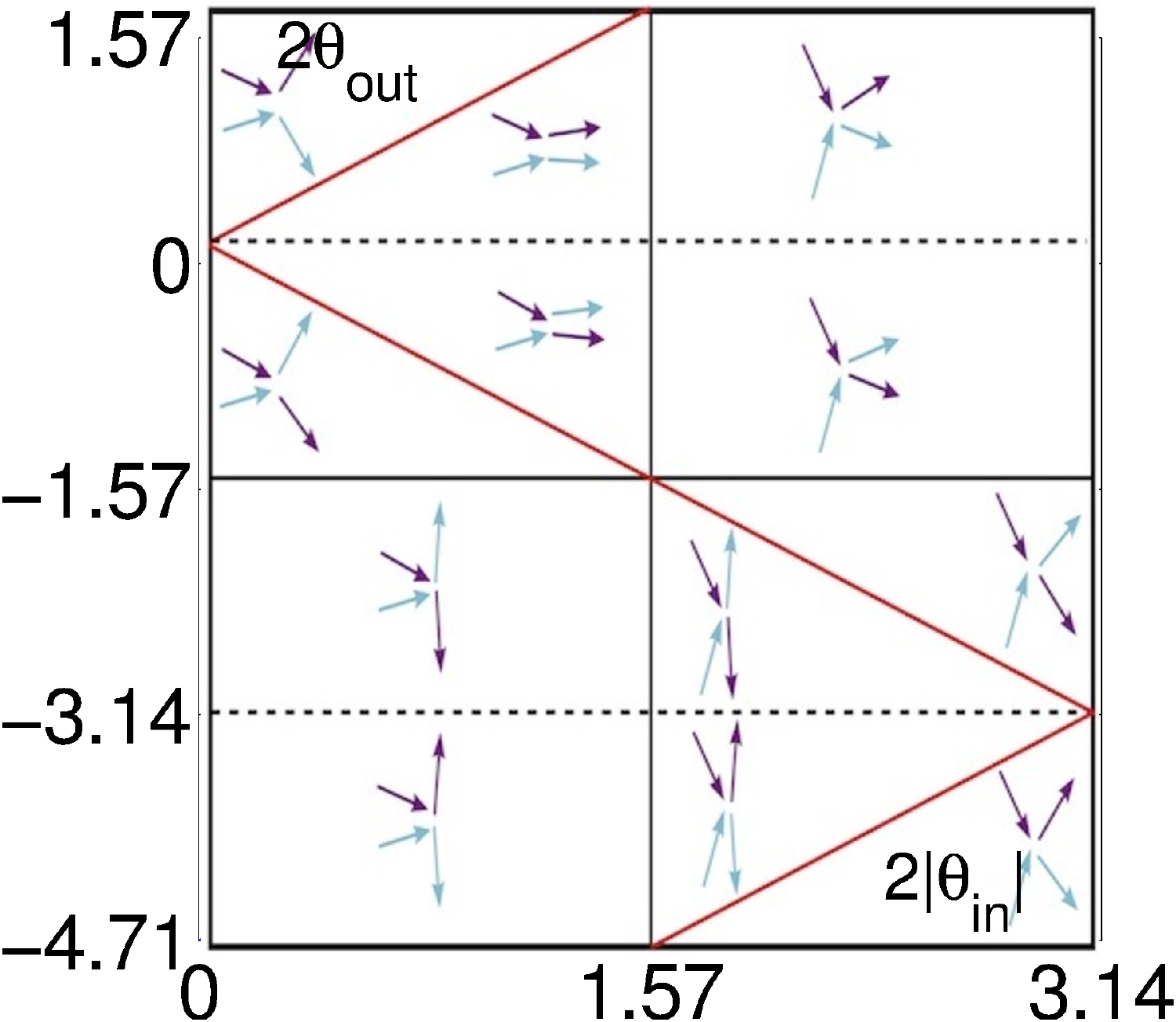}}
\hbox{\hspace{2.1cm}(a)\hspace{4cm} (b)}
}
\vspace{3mm}
\vbox{
\hbox{\hspace{2mm}
\includegraphics[clip,width=3.9cm,height=4cm]{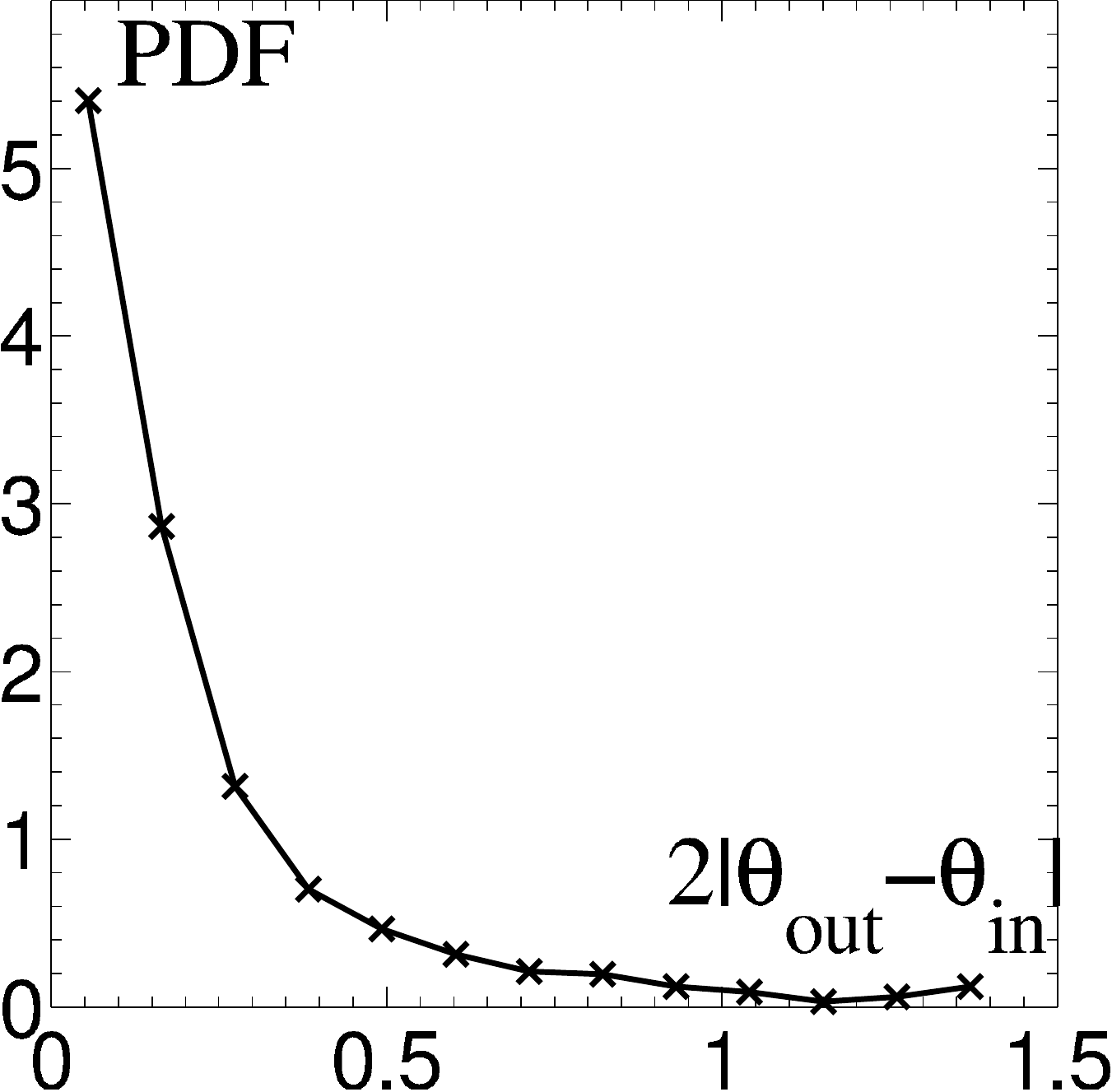}
\includegraphics[clip,width=4.4cm,height=4.1cm]{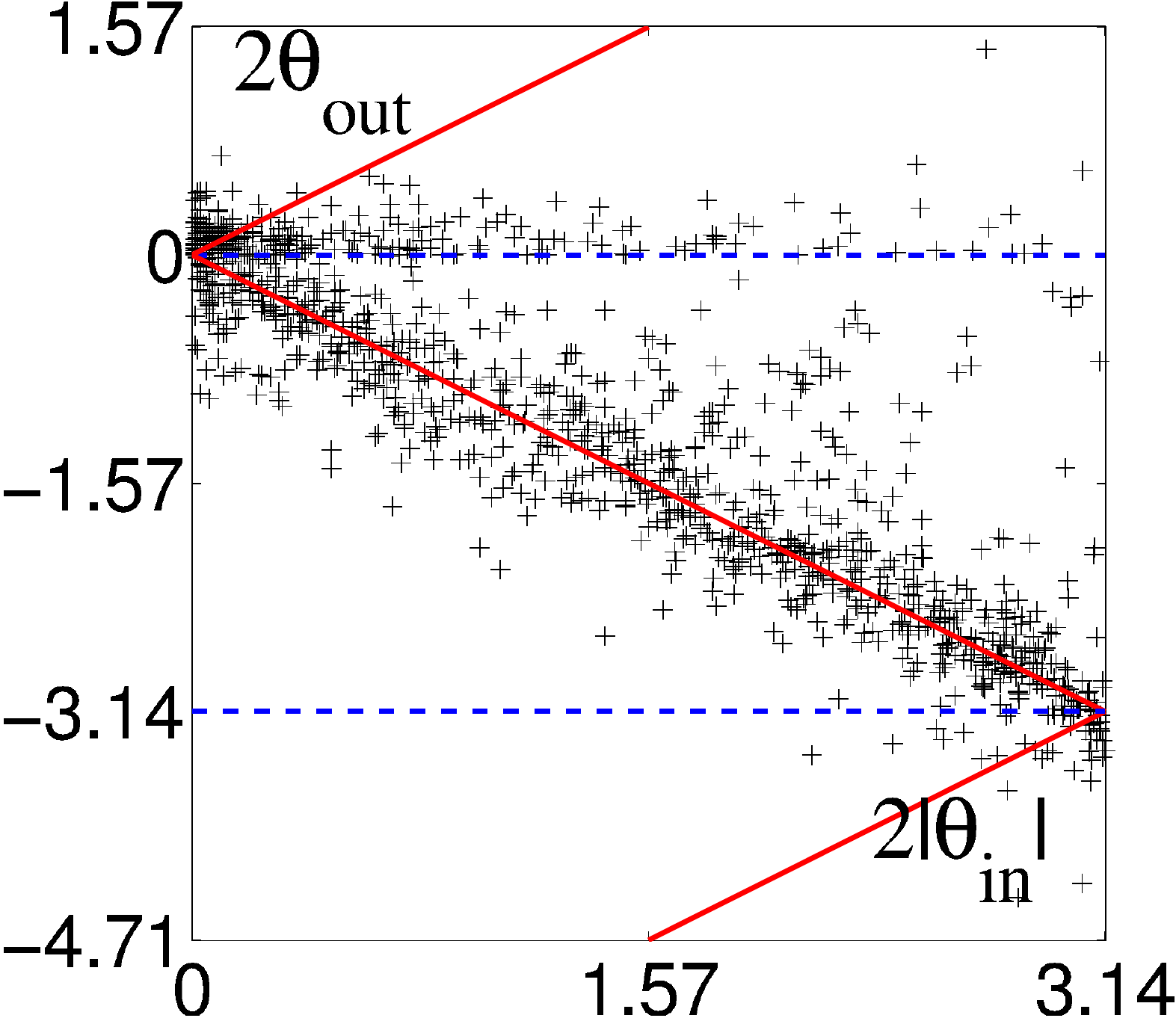}}
\hbox{\hspace{2.1cm}(c)\hspace{4cm} (d)}
}
\caption{(color online) Alignment properties of the collisions.
(a) Distribution of the deviation angle $\beta$ of the collision orientation, (see Fig.~\ref{fig:coll-stat}b) 
(b) Sketch of the collisions in the plane $(2|\theta_{in}|$,$2\theta_{out})$ 
(c) Distribution of $2|\theta_{out}-\theta_{in}|$
(d) Scatter plot of $(2|\theta_{in}|,2\theta_{out})$ characterizing the alignment properties of the collisions. 
}
\label{fig:coll-align}
\end{figure}

\subsection{Alignment properties}

In spite of their complicated space-time structure, and of the intrinsically
chaotic/noisy character of the particle dynamics, the binary collisions
possess a remarkable property: the sum of the polarities of the two particles 
changes very little between the beginning and the end of the collision.
The distribution of the angle $\beta$ characterizing this 
(see Fig.~\ref{fig:coll-stat}b) is sharply peaked around 0
(Fig.~\ref{fig:coll-align}a).

In the following, we thus assume that $\beta=0$, and neglect the dependence
of the collision statistics on the impact coefficient, so that all the information
related to alignment is encoded in the conditional probability of 
$\theta_{\rm out}$, the outgoing angle of the polarities, on $\theta_{\rm in}$, 
the incoming angle (Fig.~\ref{fig:coll-stat}b). All this information
is represented in the scatter plot of $\theta_{\rm out}$ vs $\theta_{\rm in}$
(Fig.~\ref{fig:coll-align}b,d).
A first striking fact is that most collisions actually do {\it not} change 
polarities at all: most points of the scatter plot are located near the diagonal 
$\theta_{\rm out}=-\theta_{\rm in}$; the distribution of $|\theta_{\rm out}-\theta_{\rm in}|$ is
sharply peaked near zero (Fig.~\ref{fig:coll-align}c).
In the $(\theta_{\rm in}, \theta_{\rm out})$ scatter plot, 
apart from the highly-populated diagonal region,
most of the remaining points are located near the 
$\theta_{\rm out}=0$ axis (near-perfect, Vicsek-style
alignment) or in the triangular region between this axis 
and the $\theta_{\rm out}=-\theta_{\rm in}$ diagonal
(reduced outgoing angle, but not perfect alignment) 
(see Fig.~\ref{fig:coll-align}b,d).
Anti-alignment or ``nematic'' interactions similar to that
represented in Fig.~\ref{fig:coll-stat}a (bottom)
can be seen, but they remain rare. 

Thus, all in all, about 70\%
of binary collisions amount to no  effective interaction. Among the 
30\% remaining ones, most are strongly aligning, with a few outliers. 
So far, our system seems to be rather close to the Vicsek model: 
our particles can be considered, on some coarse-grained timescale, 
to be moving at  constant speed. Via the shaking amplitude $\Gamma$,
we have good control on the effective 
rotational diffusion constant $D_\theta$. And binary collisions, 
when they significantly alter the orientation of particles, are clearly aligning.

This said, it is still by no means clear, in the higher density regimes where 
collective motion could arise, that any of these observations 
is really important. Indeed, one
expects there short mean free paths, and very complicated, 
possibly unending collision events
involving more than two particles, a usual feature of dense 
collections of finite-size granular
particles made even more acute here by the fact that binary 
collisions are spatio-temporally extended events.

\section{Collective motion}
\label{collective}

We now turn to the study of the collective dynamics of our vibrated polar disks.
We first discuss interactions between our particles and the lateral walls of
our apparatus, as well as the choice of the size and shape of the horizontal
domain in which they are evolving. 

\subsection{Domain size and shape, collisions with boundary}

\begin{figure}[!t] 
\centering
\includegraphics[width=8.4cm]{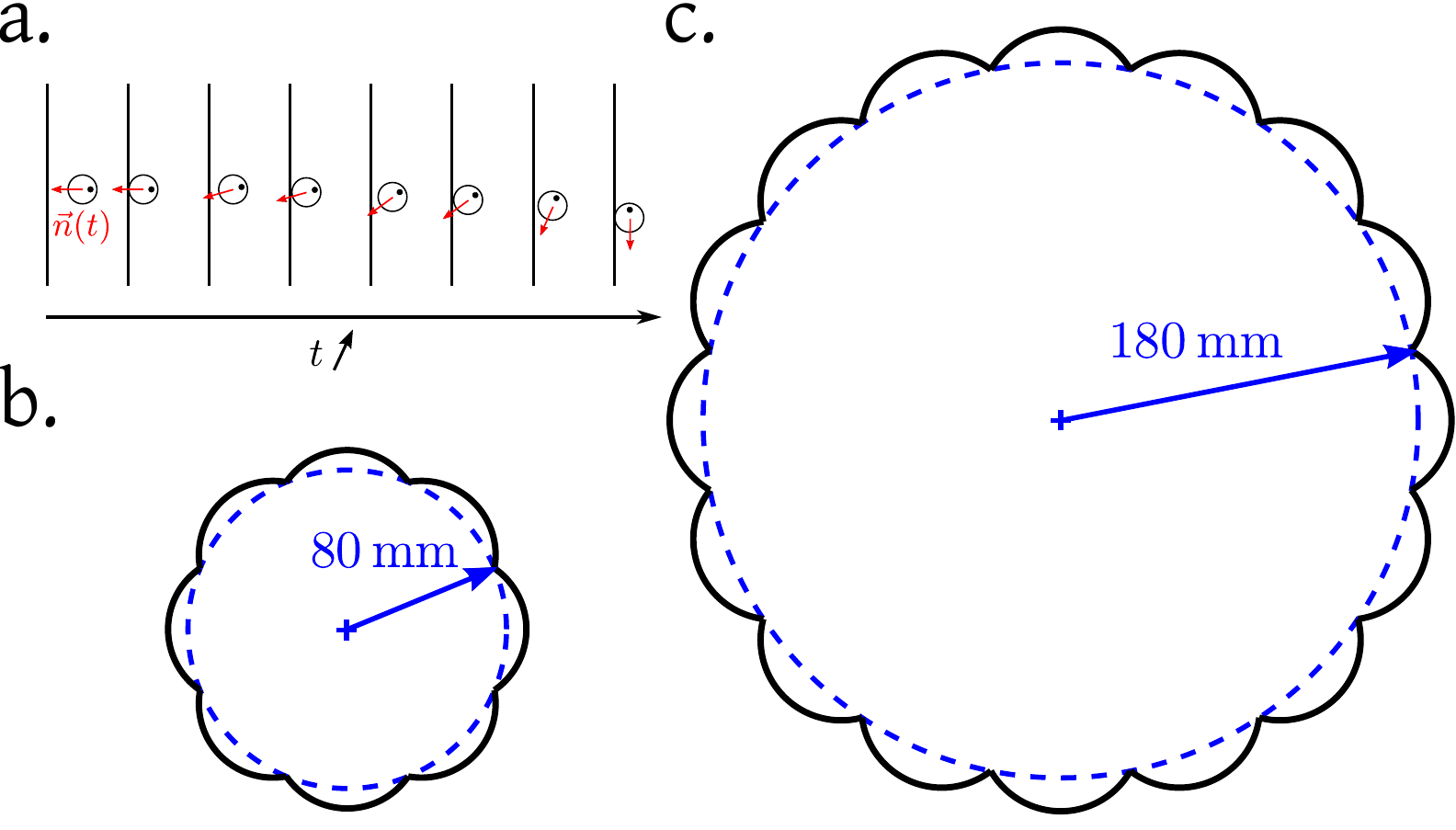}
\caption{(color online) Boundary conditions:
(a) Sketch of the mechanism responsible for the accumulation of self-propelled disks at the walls. 
(b) Small system: internal radius (dashed circle) is $20$ and the arcs radii are $9.3$. 
(c) Large system: internal radius (dashed circle) is $45$ mm and the arcs radii are $10.125$.}
\label{fig:bcs}
\end{figure}

When one of our particles hits a wall head-on, it bounces on it
repeatedly, gradually turning after each hit, and eventually escapes, typically
along the wall (see a sketch in Fig.~\ref{fig:bcs}a). Thus, in a many-particle 
system, one expects an accumulation of particles near the wall, be it flat or
circular. This is a common problem when dealing with self-propelled
objects, as already experienced, e.g., by Kudrolli {\it et al.}
\cite{kudrolli2008swarming}.  In the ``real world'',  one cannot easily rely on the
periodic boundary conditions available in numerical studies,
short of having particles move on some closed surface like a sphere or a torus.

In the present case, we designed ``flower-shaped'' domains (Fig.~\ref{fig:bcs}bc)
to help particles escape from the wall. Crawling along a ``petal'', they are likely 
to be reinjected in the bulk when reaching the cusp point marking
the limit with the next petal: each arc behaves as a launchpad.
Two different flower-shaped arenas have been used in the following, 
one of internal diameter 40 and 8 petals (the ``small'' system) 
the other one of internal diameter 90 and 16 petals (the ``large'' system).
The ``petal'' size was {\it not} scaled with the internal diameter, but was
kept roughly constant: their arc radii are 9.3 for the small system, and 10.125 for the large one.
Therefore, only the general available area increases when changing 
from the small to the large system, not the length-scale used to re-inject
particles toward the center.

\subsection{Packing fraction and region of interest}

In the vast majority of models for collective motion, the ordered, 
collectively-moving phase is reached by either increasing the persistence
length of the single particle walk (e.g. by reducing the angular noise strength)
or increasing the density of particles (which make interactions frequent enough
to sustain the memory of the last aligning collision until the next event).

In our system, we have seen that the persistence length is controlled rather
nicely by the vibration amplitude $\Gamma$. 
Clearly, this is the easiest control parameter to use here, since changing 
density would be tedious. But we already showed that the
polar disks in our system can only move coherently
on a rather restricted range of $\Gamma$ values
(in particular propulsion deteriorates quickly when $\Gamma<2.7$). 
This limit, in turn, imposes that we work at the ``right'' density of particles
so that the transition to collective motion and the ordered phase be observable.
If the density is too low, our system cannot order, even for $\Gamma=2.7$. 
If it is too high, on the other hand, one is bound to run into the problem of
glassy dynamics, crystallization and jamming: 
because of their finite size, our polar particles flock in packed clusters.
Jamming of active particles is an interesting problem (on which research
has just started \cite{angelini2011glass,henkes2011active}), but it is beyond the
``simpler'', Vicsek-style problem of collective motion of point particles
considered here.

\begin{figure*}[t!]
\centering
\hbox{
\vbox{
	\hbox{\hspace{1mm}
		  \includegraphics[clip,width=5.4cm,height=5.5cm]{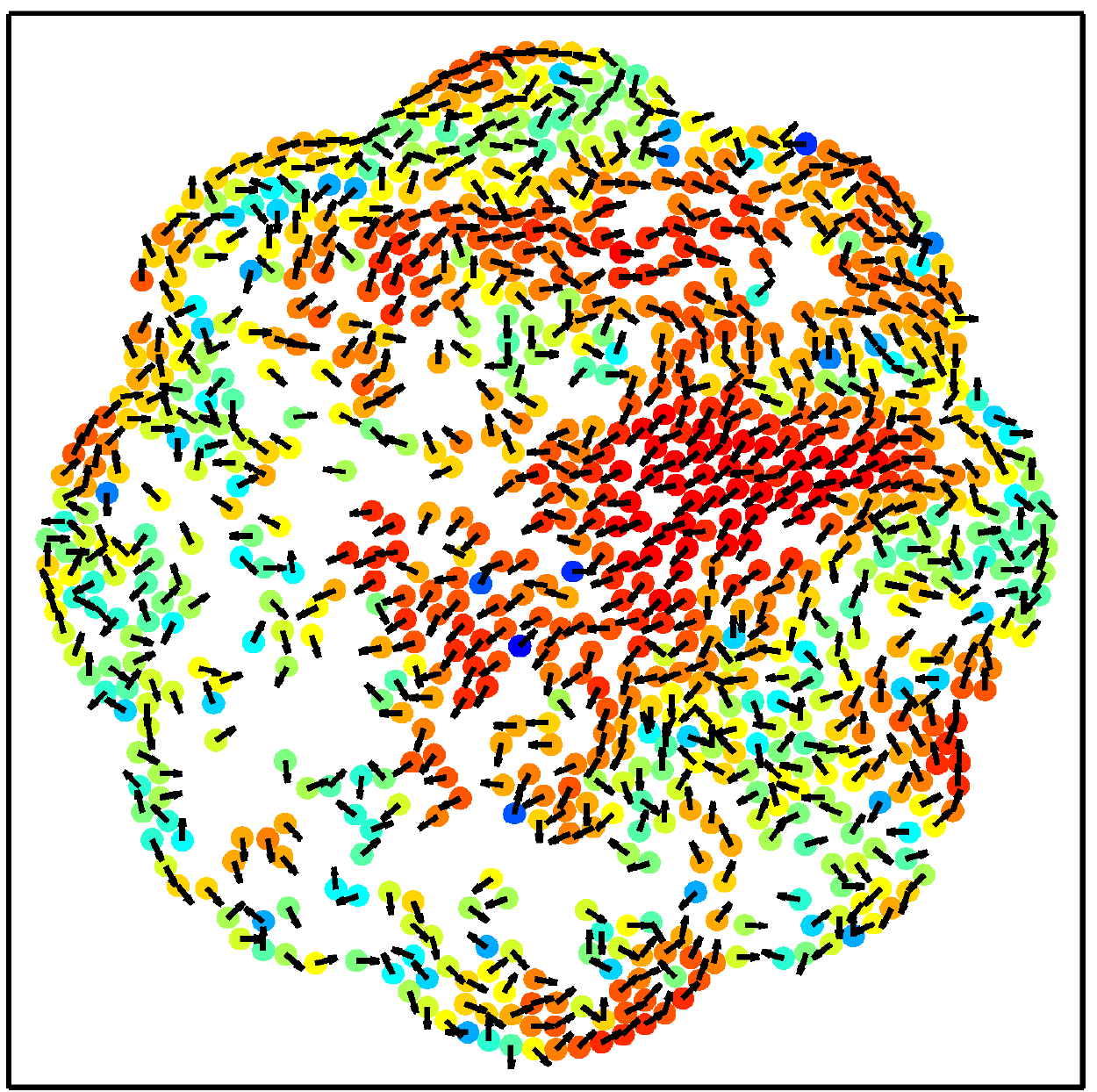}
		  \includegraphics[clip,width=5.5cm,height=5.5cm]{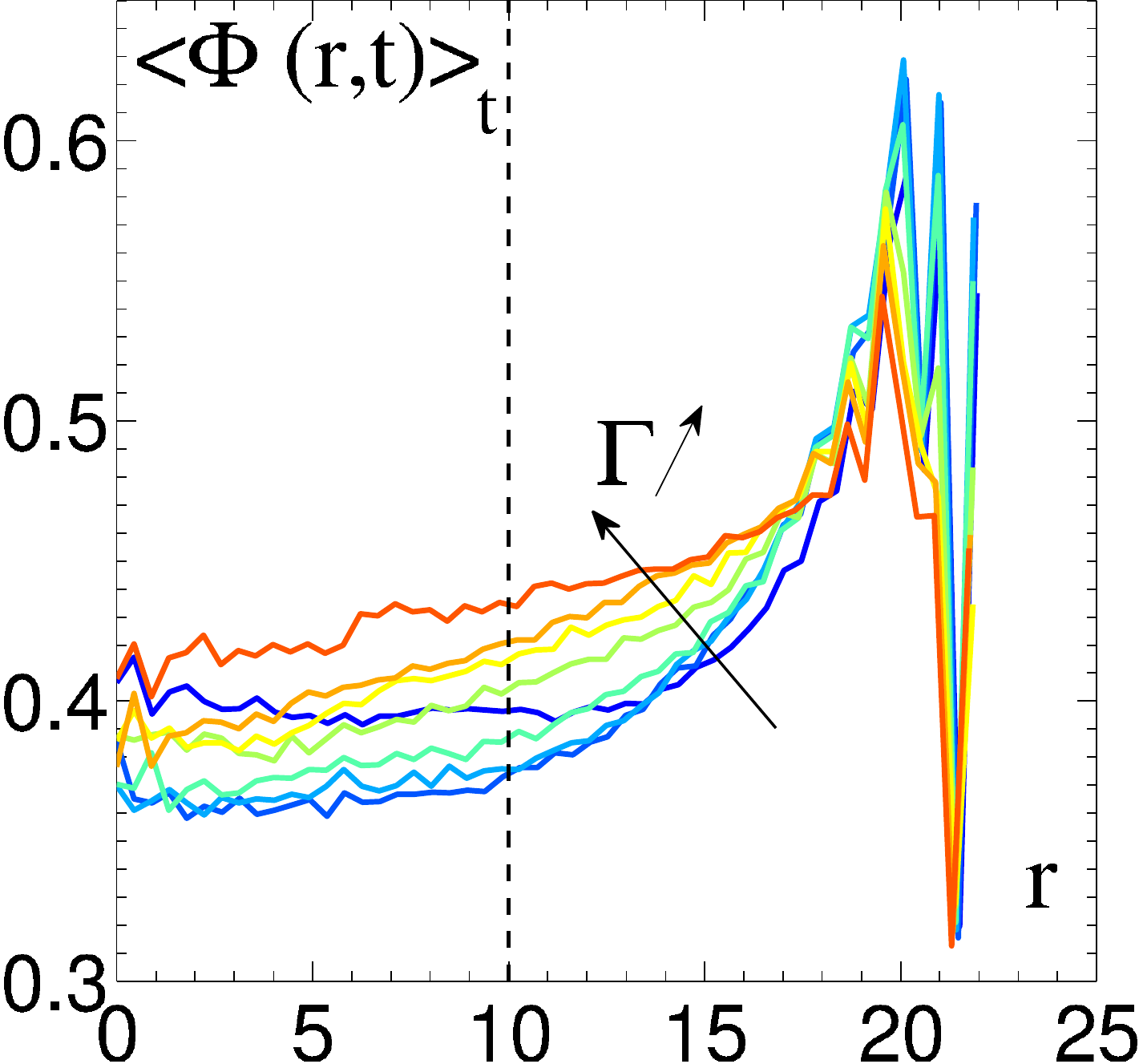}}
	\hbox{\hspace{2.7cm}(a)\hspace{5cm} (b)}
	\hbox{\includegraphics[clip,width=5.5cm,height=5.5cm]{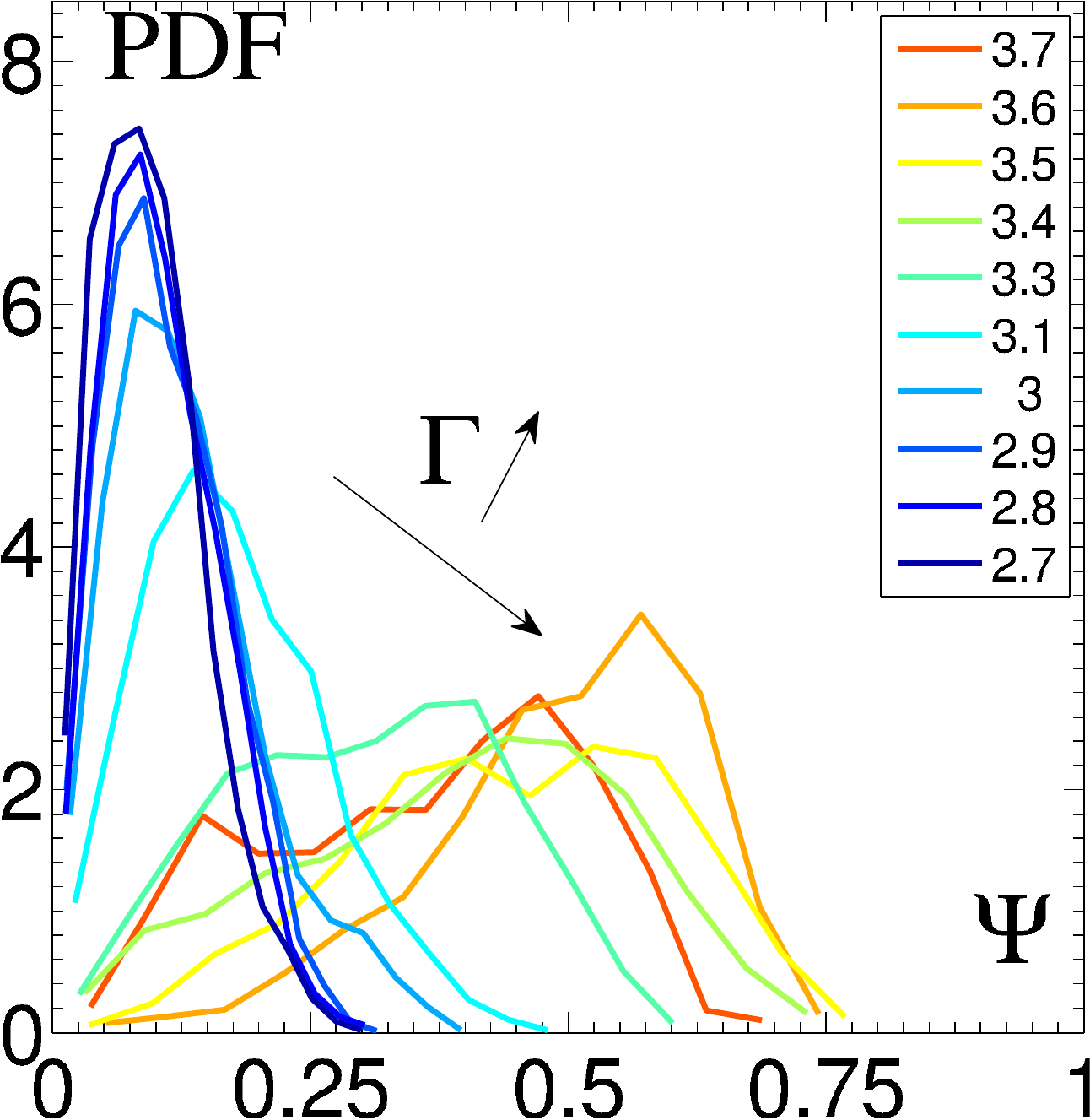}
		  \hspace{0.5mm}
                    \includegraphics[clip,width=5.5cm,height=5.5cm]{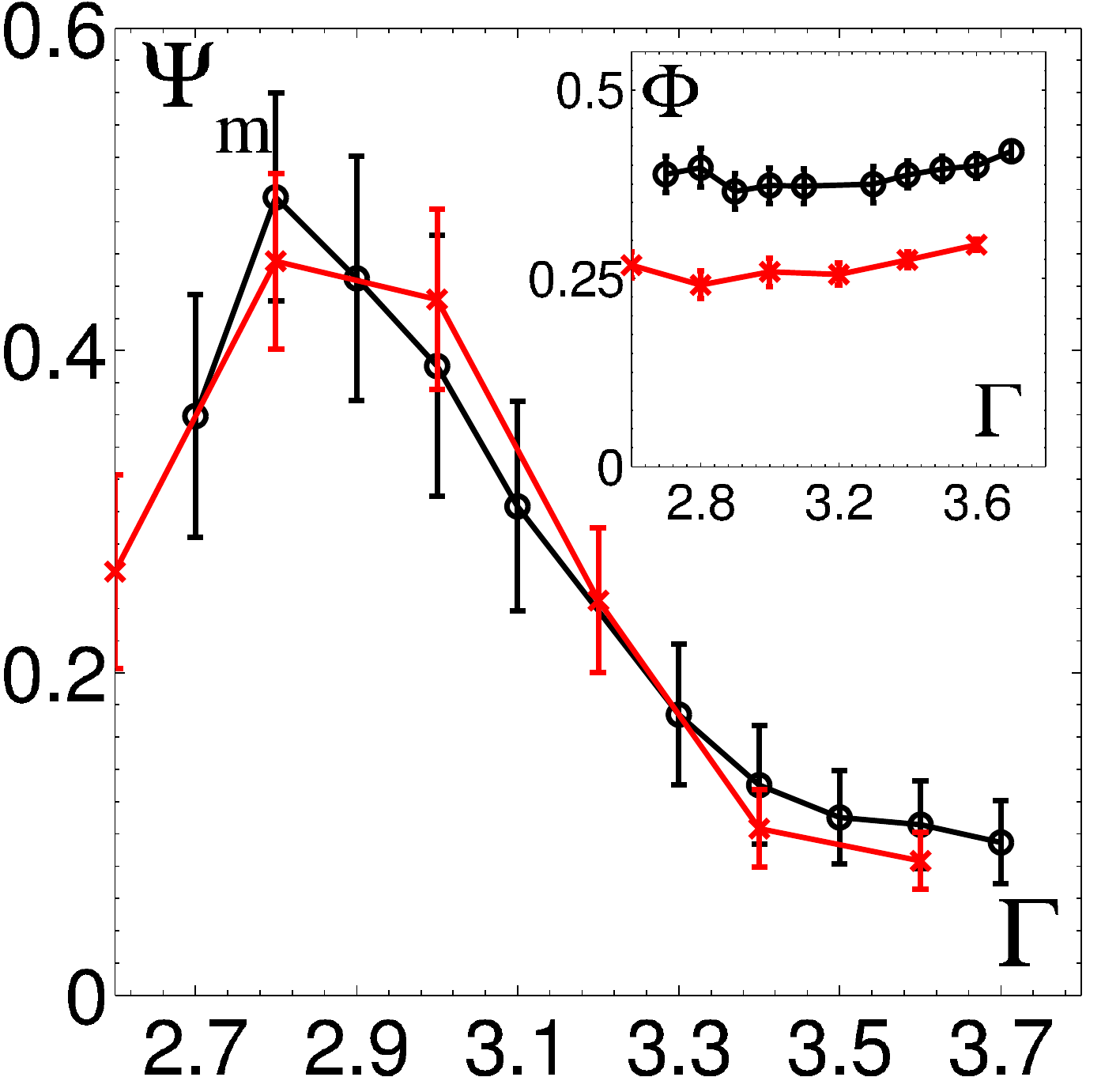}}
	\hbox{\hspace{2.7cm}(c)\hspace{5cm} (d)}
	}
\hspace{-6cm}
\vbox{\includegraphics[clip,width=5.7cm,height=2.85cm]{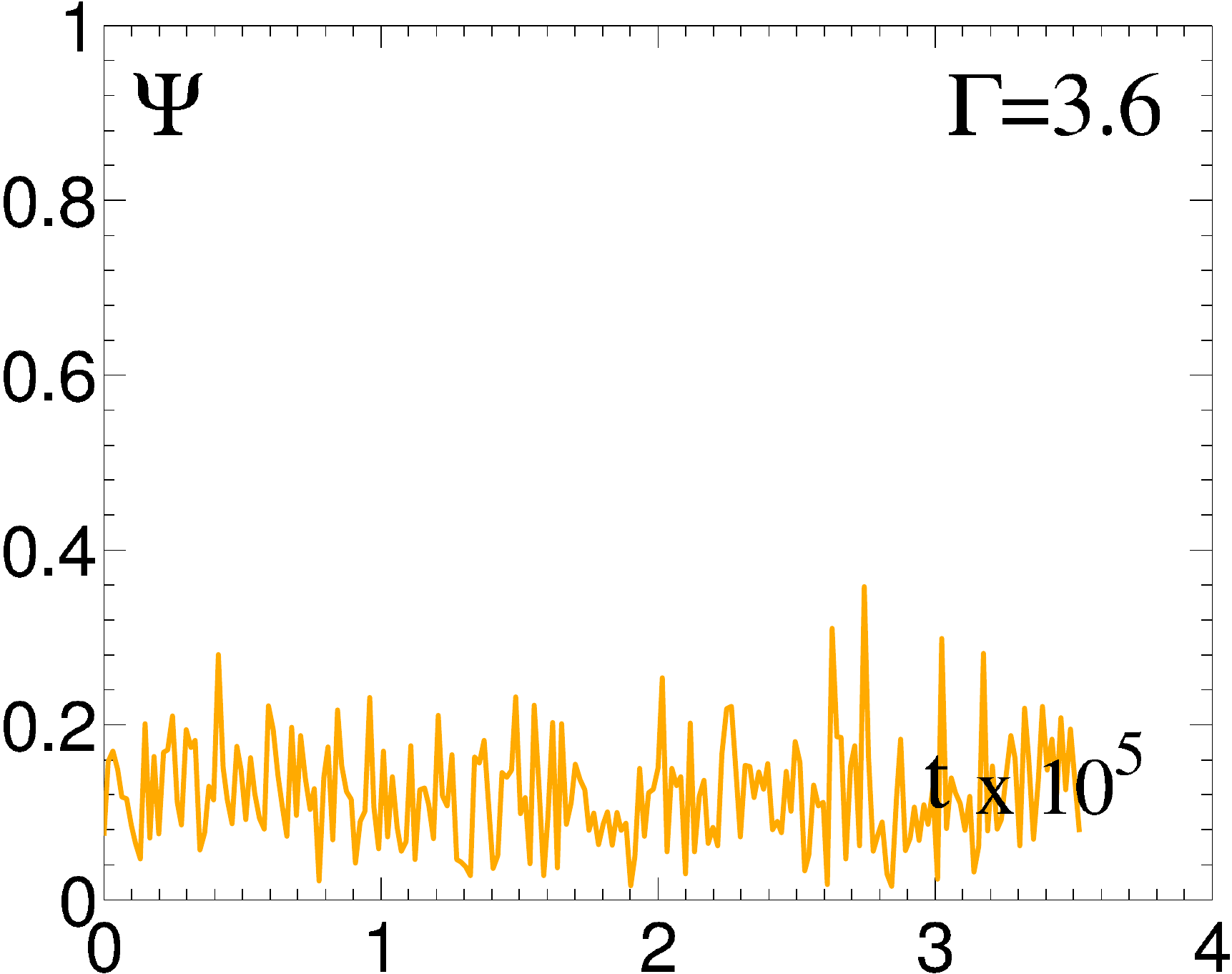}\\
	 \includegraphics[clip,width=5.7cm,height=2.85cm]{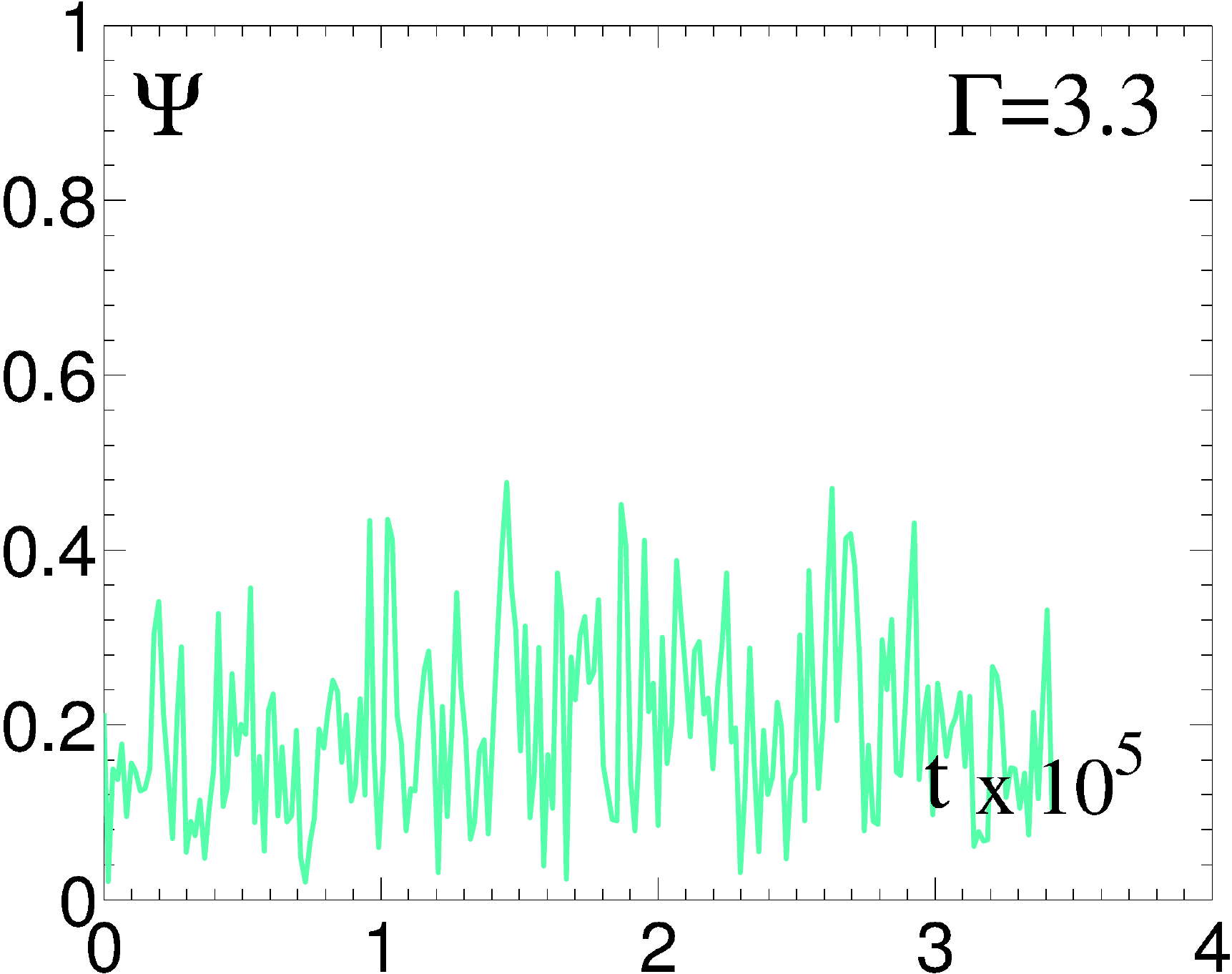}\\
	 \includegraphics[clip,width=5.7cm,height=2.85cm]{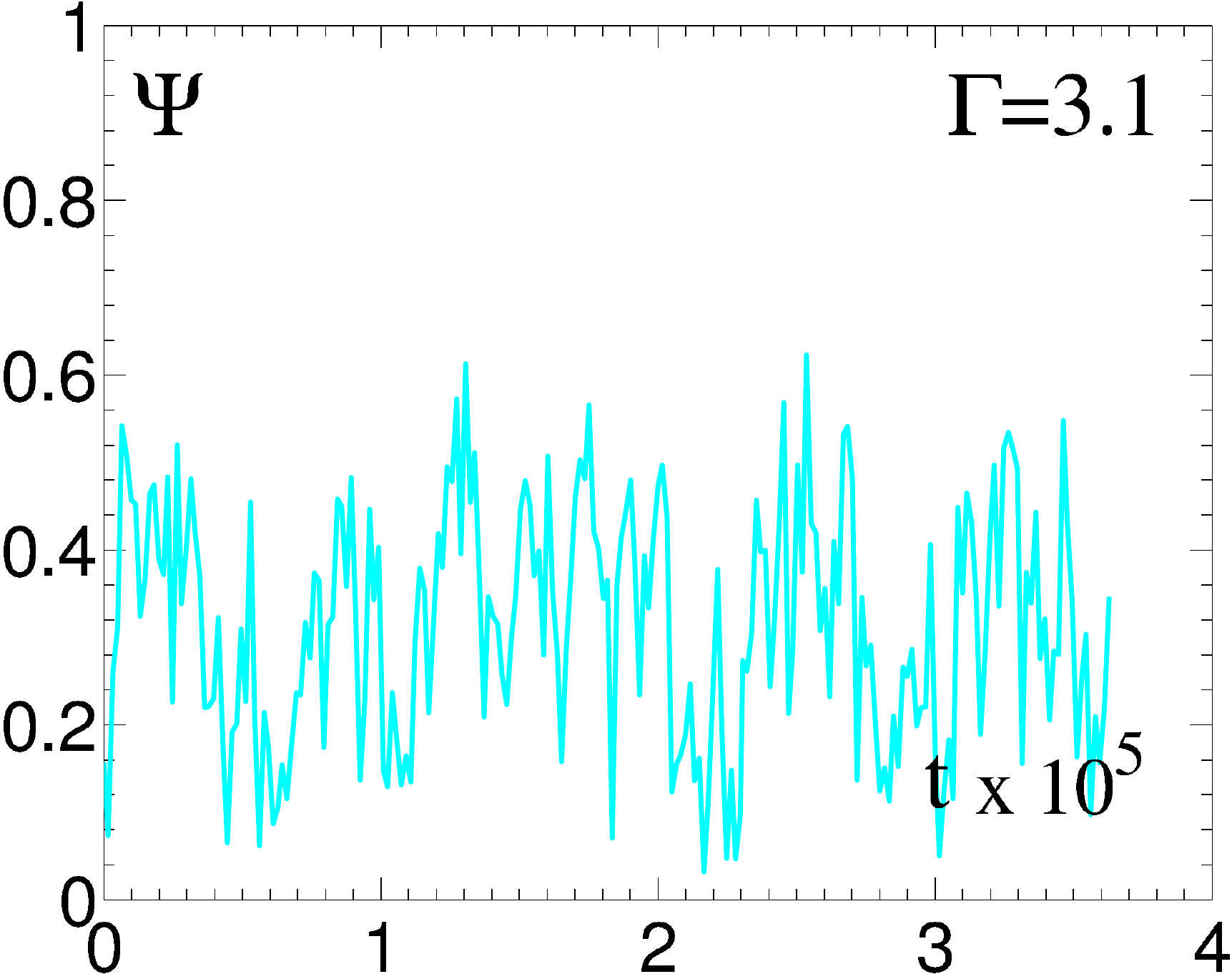}\\
	 \includegraphics[clip,width=5.7cm,height=2.85cm]{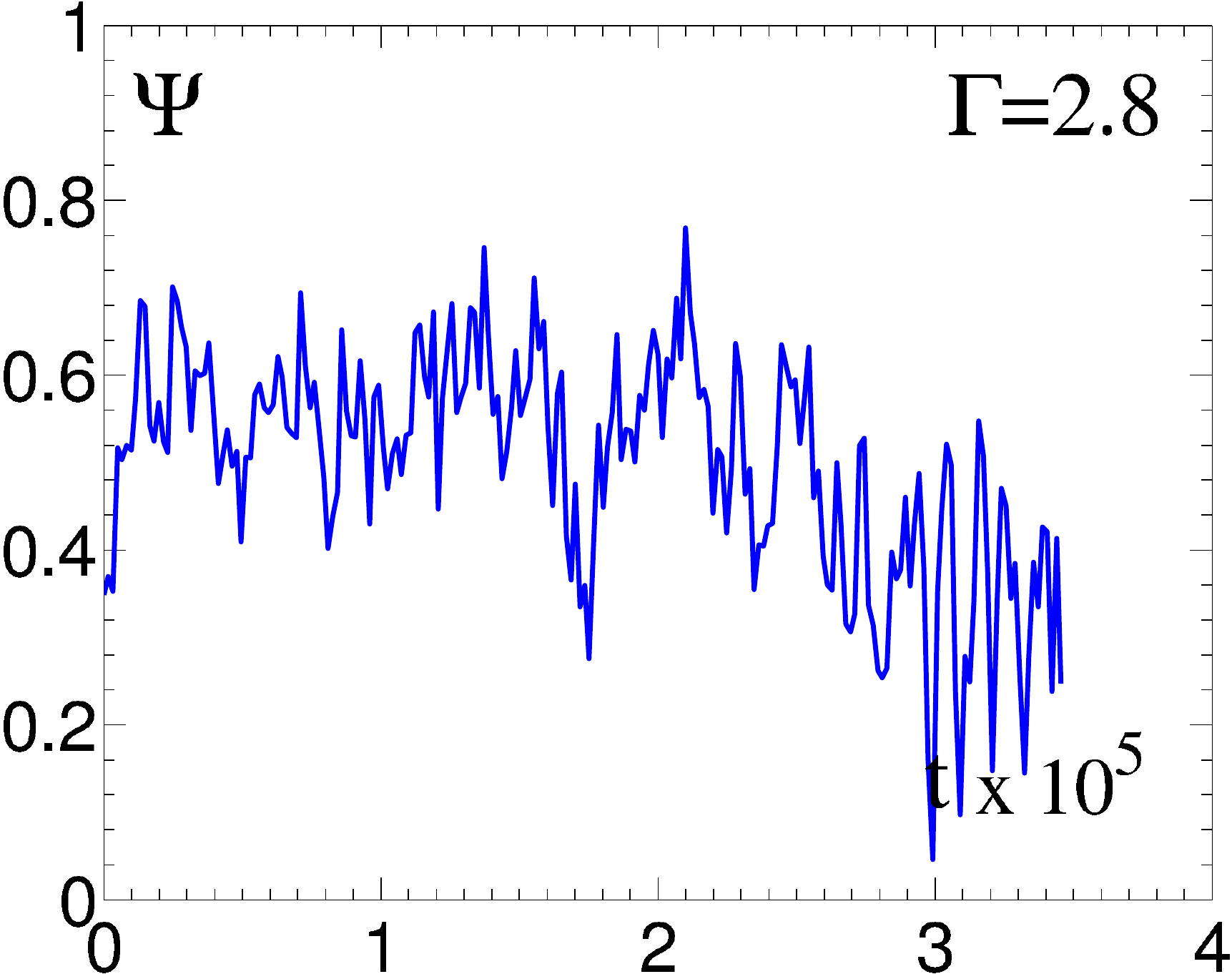}\\
	 (e)
	}
}
\caption{(color online) Emergence of collective motion:
(a) Snapshot taken in a regime with long-range collective motion. The color code is red (reps. blue) for perfectly aligned (reps. anti-aligned) neighbors. 
(b) Time-averaged packing fraction $\langle\phi(r,t) \rangle_t$ versus the distance $r$ to the center of the vibrating plate for $\Gamma\in[2.7-3.7]$. The size of the region of interest (ROI) is chosen in order to ensure uniformity of the packing fraction inside the ROI and indicated by the vertical dashed line. The ROI in the small system has a radius of 10 particles diameters. We also checked that the packing fraction inside the ROI is stationary. (Same color code as in (c))
(c) Distribution of the order parameter $\Psi$ for $\Gamma\in[2.7-3.7]$, (color code as indicated in the legend).
(d) Most probable value $\Psi_m$ of the order parameter as a function of $\Gamma$ for both the smaller experiment {\sf Ea} (black $\circ$) and the larger one, {\sf Eb}, (red $+$). Inset: $\Phi_{{\rm ROI}} $ as a function of $\Gamma$ for the same experiments.
(e) Temporal evolution of the order parameter $\Psi(t)$ for four different vibration amplitudes $\Gamma=3.6, 3.3, 3.1, 2.8$. }
\label{fig:coll-motion}
\end{figure*}

Thus, that collective motion is observable in our system is not guaranteed.
We found, by trial and error, that nominal packing fractions $\phi$
(the total area covered by the particles in the system divided by the domain area) 
in the range $\phi\in [0.4,0.6]$ give the best results. For instance,
890 particles in our ``small'' domain ($\phi=0.47$) vibrated at $\Gamma=2.7$
move collectively on large scales in a spectacular way 
(Fig.~\ref{fig:coll-motion}a and Supplementary movie \cite{epaps3}).

In fact, the issues discussed above are further complicated by the effect of
boundaries. Even if our flower-shaped domains help solve the problem of
particle accumulation along the walls, they do not prevent it entirely,
and we do observe, on average, higher densities near the wall
(Fig.~\ref{fig:coll-motion}b). This forces us to define a 
``region of interest'' (ROI) over which the observed 
time-averaged density remains approximately constant. 
From graphs like that of Fig.~\ref{fig:coll-motion}b, we see that 
the time-averaged density is constant to a few percent 
if the ROI is chosen to be of diameter 20, i.e. half the nominal diameter of our small system. 
For the range of packing fractions we will discuss below, we could safely use 
the same ROI size independently of the nominal packing fraction.
Finally, going from our small domain (diameter 40) to the large one (diameter 90),
we find that the ROI roughly doubles (from diameter 20 to 40)
if one keeps the requirement of a few percent variation of the observed
time-averaged density.  
Inside the ROI, the well-defined averaged packing fraction $\phi_{\rm ROI}$
is not only significantly smaller than the nominal packing fraction $\phi$, 
but it decreases slowly with $\Gamma$ (see inset of Fig~\ref{fig:coll-motion}d). 
(Note that these are opposite effects: decreasing $\Gamma$ should increase
order since this reduces $D_\theta$, but lower $\phi_{\rm ROI}$ values should decrease it.)
For the 890 particles in our small system,
we observe that $\phi_{\rm ROI} \in [0.36,0.42]$ for $\Gamma\in [2.7,3.7]$
(Fig.~\ref{fig:coll-motion}b).

With these remarks in mind, we now present results from three sets of experiments
performed by varying $\Gamma$ at fixed ``geometrical conditions'' (Table~\ref{table}).

\begin{table}[ht]
\caption{\label{table}
Main characteristics of the sets of experiments described.}
\begin{ruledtabular}
\begin{tabular}{ccccc}
Experiment & Arena used & \# of particles & $\phi$ & typical $\phi_{\rm ROI}$
\\ \hline
{\sf Ea} & small & 890 & 0.47 & 0.39 \\
{\sf Eb} & large & 3830 & 0.42 & 0.26 \\
{\sf Ec} & small & 1090 & 0.58 & 0.19 
\end{tabular}
\end{ruledtabular}
\end{table}

\subsection{Onset of collective motion}

In experiments {\sf Ea} and  {\sf Eb}, decreasing the vibration amplitude $\Gamma$
brings the system from a disordered state to regimes where particles, in 
large regions of the system of the order of the ROI diameter, move in
approximately the same direction
(Fig.~\ref{fig:coll-motion}a and Supplementary movie\cite{epaps3}).

Different order parameters can be defined to quantify the degree of orientational
order. The average polarity is easy to measure, but we preferred to use the average 
orientation of the displacement of each particle over a time $\tau=50$,
taken as a proxy for velocity. Although both choices yield essentially the
same results, the latter one is more directly related to the transition to collective {\it motion}.
Below, $\Psi(t)$ in fact represents the modulus of this velocity-based
order parameter, properly normalized to be 1 for perfectly aligned particles.

\begin{figure*}[t!]
\vbox{
\hbox{
\hspace{5mm}
\includegraphics[clip,width=5.5cm,height=5.5cm]{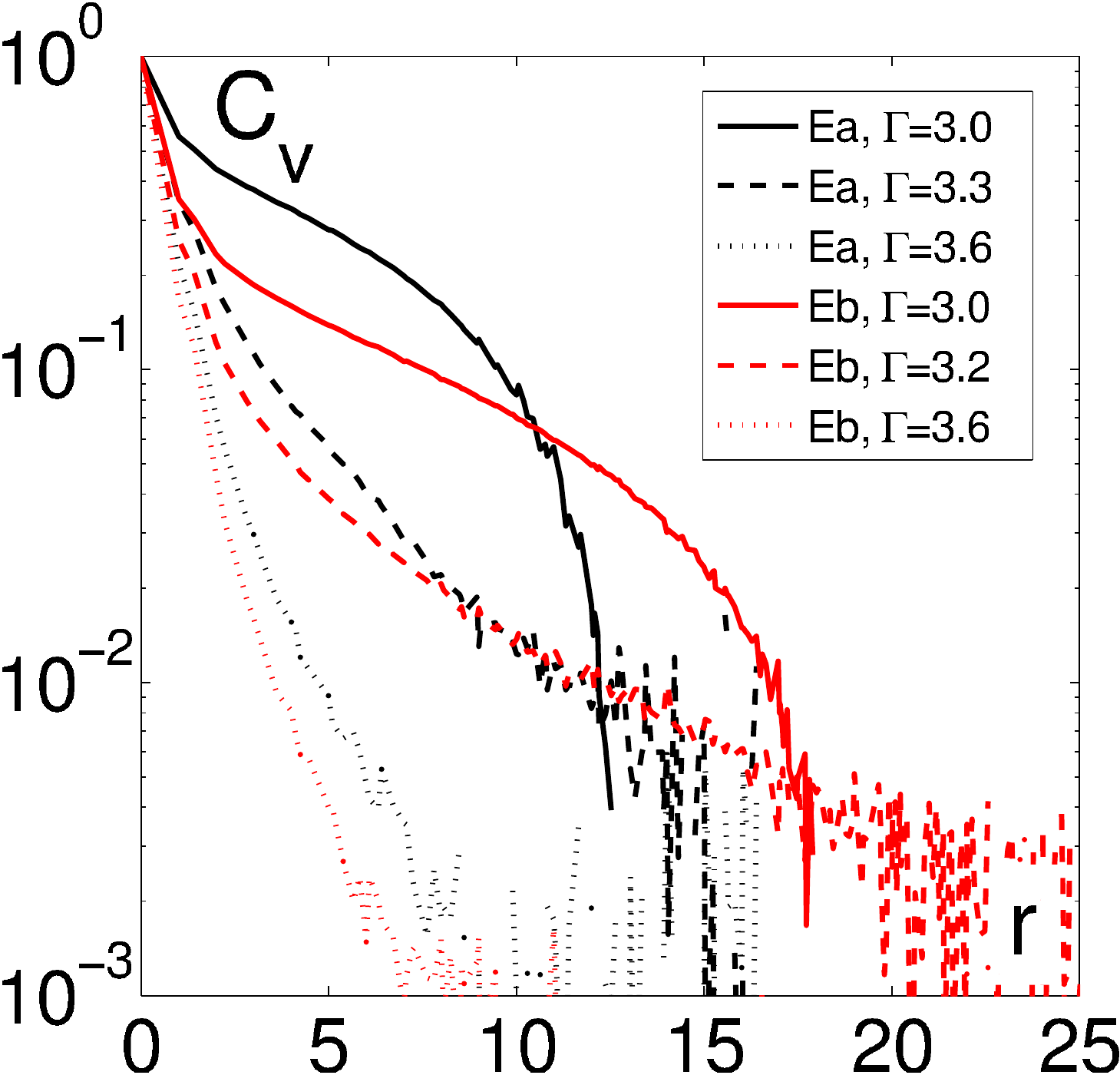}
\includegraphics[clip,width=5.5cm,height=5.5cm]{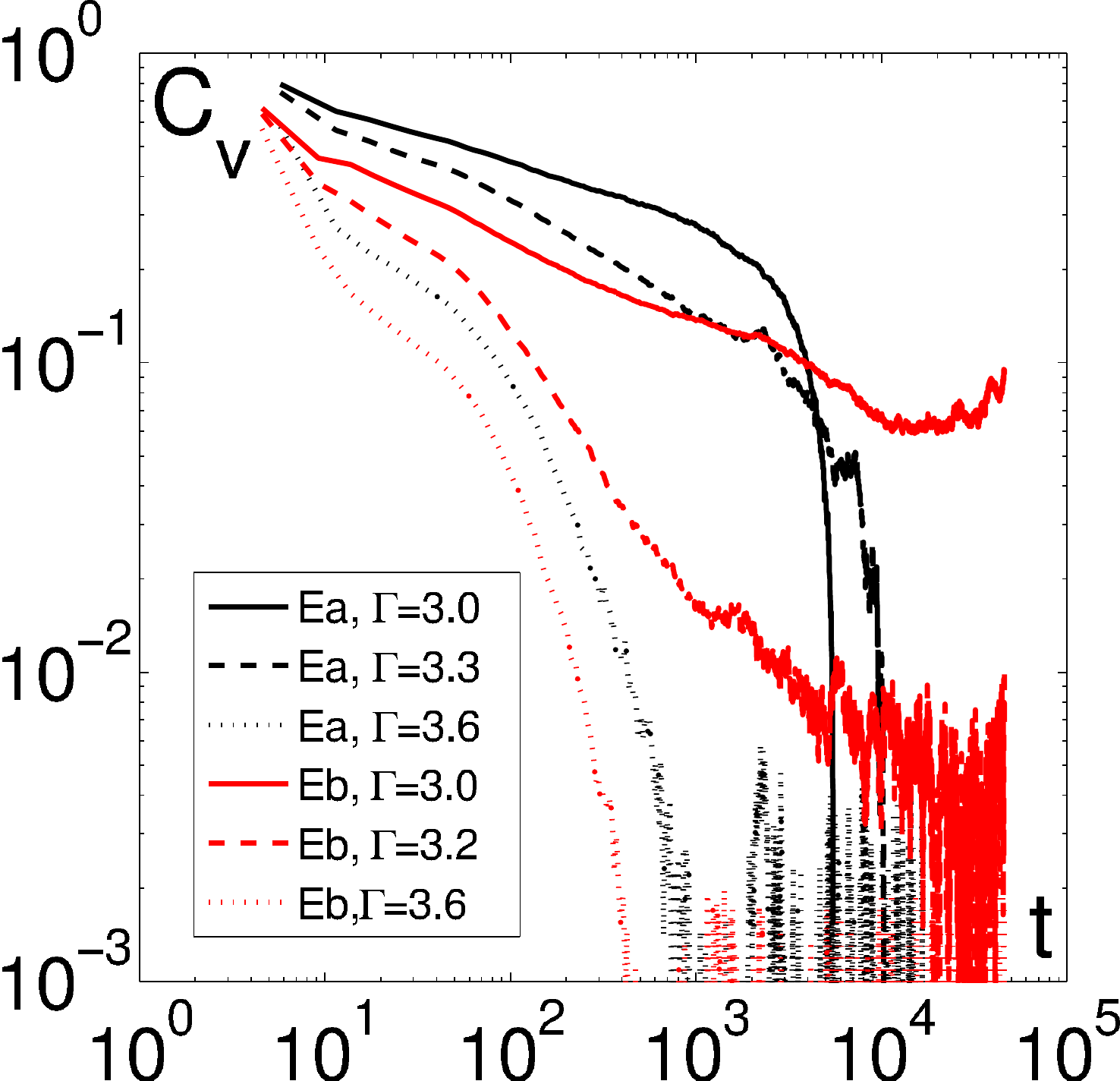} 
\includegraphics[clip,width=5.5cm,height=5.5cm]{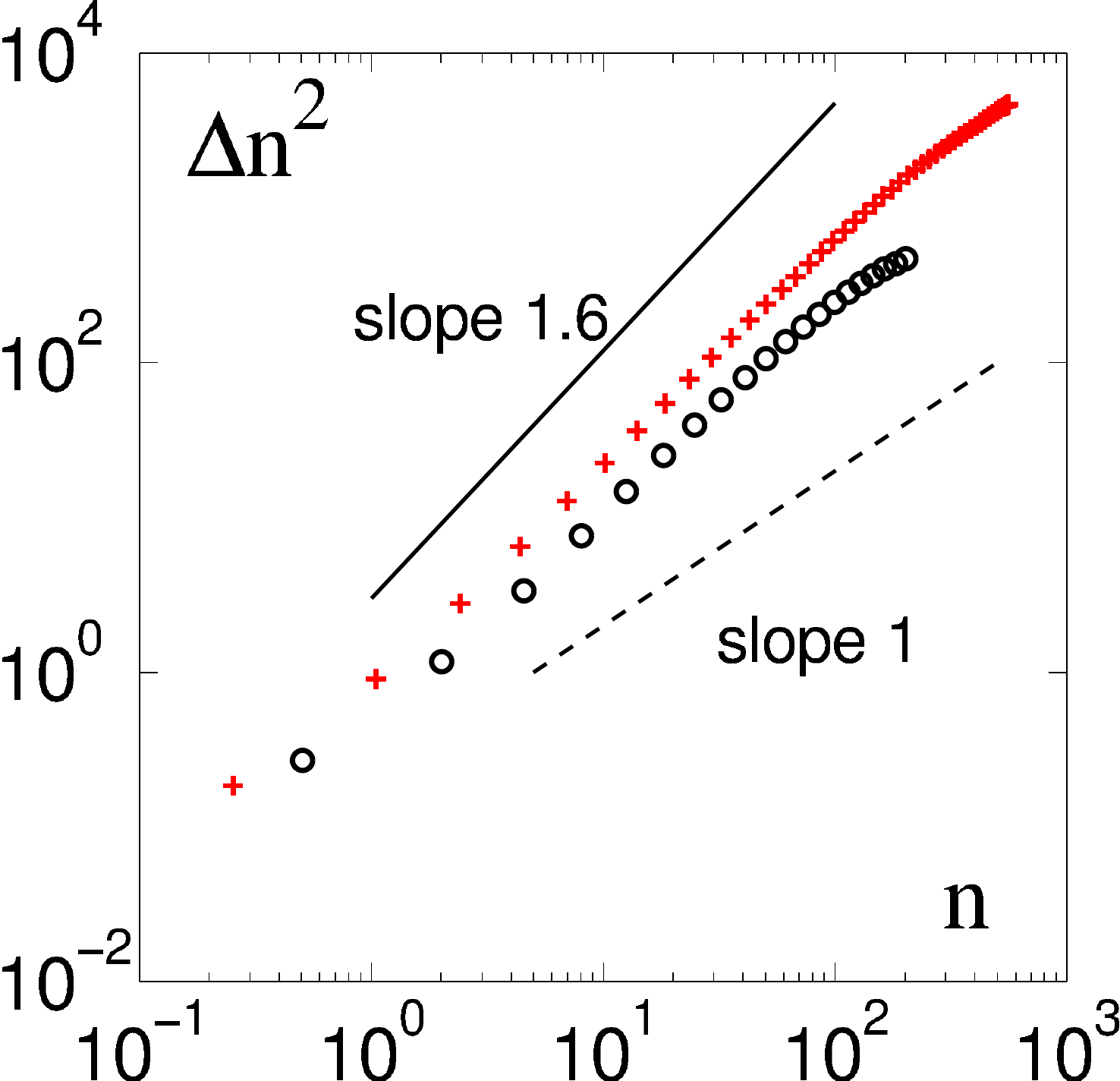} }
\hbox{\hspace{3.5cm}(a)\hspace{5.2cm}(b)\hspace{5.2cm} (c)}
}
\caption{(color online) Finite-size effects.
(a) Spatial correlation $C_v(r)$ of the Eulerian velocity field $\vec{v}(\vec{r},t)$ for $\Gamma = 2.8$ for the small (black curves) and the large (red curves) systems at three different vibration amplitudes $\Gamma$ as indicated in the legend.
(b) Temporal correlation $C_v(t)$ of $\vec{v}(\vec{r},t)$ for $\Gamma = 2.8$ for the small (black curves) and large (red curves) systems at three different vibration amplitude $\Gamma$ as indicated in the legend.
(c) Fluctuations $\Delta N^2$ of the number of particles as a function of the average number of particles $N$ in boxes of increasing size for the small (black $\circ$) and large (red $+$) systems. 
} 
\label{fig:finite-size}
\end{figure*}

Time series of $\Psi(t)$ show that decreasing $\Gamma$ from 3.7 to 2.7, the system
reaches more and more ordered states, albeit never perfectly ordered ones as 
expected in a finite system with hard walls (Fig.~\ref{fig:coll-motion}e). 
The most ordered regimes exhibit very large fluctuations 
of the order parameter over very long timescales as shown also in the
corresponding histograms of values taken by $\Psi(t)$
(Fig.~\ref{fig:coll-motion}c).

As a matter of fact, our limited range of usable $\Gamma$ values did not allow 
to reach regimes well inside the ordered phase, as shown, e.g., by the variation
with $\Gamma$ of $\Psi_m$, the most probable value of the order parameter
(Fig.~\ref{fig:coll-motion}d). In both experiments {\sf Ea} and  {\sf Eb},
 $\Psi_m$ reaches values of the order of 0.5 for $\Gamma=2.8$, 
but starts falling back for smaller $\Gamma$ values. In the experiment on the 
large system {\sf Eb}, the rise a $\Psi_m$ seems steeper than in experiment
 {\sf Ea}, but the two curves are in fact difficult to compare given that the
two experiments possess different averaged packing fractions in their ROI.

Thus it is not clear from the variation of the order parameter alone
whether our system, in its most ordered regime ($\Gamma=2.8$),
has reached the ordered phase or is still influenced by the transitional region. 
A direct measurement of the spatial and temporal correlation functions of the 
Eulerian velocity field suggests, though, that  the ordered phase 
is indeed reached. The Eulerian velocity field is here defined on a grid with a
square mesh of size 1x1. At each time step the local Eulerian velocity is
the average of the instantaneous velocity of the particles, the center of which is
inside one element of this mesh.
For small-enough $\Gamma$ values,  correlations in space remain strong
up to scales where the correlation  function is cut-off by the system size,
signaling that order is established on scales as large as possible (Fig.~\ref{fig:finite-size}a). 
Similarly, the temporal autocorrelation function decays very slowly (logarithmically?)
for small $\Gamma$ values and is only cut-off at very large timescales (Fig.~\ref{fig:finite-size}b).

\subsection{Giant number fluctuations}

The orientationally-ordered phases of  active matter systems are endowed with
specific yet universal properties such as  long-range correlations, superdiffusion, and strong density fluctuations \cite{ramaswamy2010mechanics,toner2005hydrodynamics}.
The latter, which go under the vocable ``giant number fluctuations'', are
the better known, probably because they are rather easy to measure: one simply 
records, in subsystems containing on average $n$ particles, the variance 
$\Delta n^2$ of the fluctuations in time of this number. For normal, equilibrium,
systems, and/or for non-interacting particles 
$\Delta n^2$ scales like $n$ at large $n$ (as long as, of course, $n\ll N$, the total
number of particles in the system). But for orientationally-ordered active systems,
one expects $\Delta n^2 \sim n^\gamma$ with $\gamma>1$.

A generic simple argument by Ramaswamy {\it et al.} \cite{ramaswamy2003active}, links a 
$q^{-2}$ divergence in the structure factor, calculated in some linear, mean-field 
approximation, to $\gamma=1/2+1/d$ where $d$ is the space dimension. This is 
expected to hold for so-called ``active nematics'', and was indeed measured
there \cite{chate2006simple}, but not for collections of polar particles as considered here.
In this case, fluctuations must be taken into account, which has been
done in the RG calculation by Toner and Tu \cite{toner1998flocks}. This calculation, 
which relies on assumptions \cite{toner2009fast}, 
predicts $\gamma=8/5$ for $d=2$, a number consistent 
with numerical results obtained on the Vicsek model \cite{chate2008collective}.

\begin{figure*}[t!]
\centering
\vbox{
\hbox{\hspace{1cm}
	\includegraphics[clip,width=8cm,height=5cm]{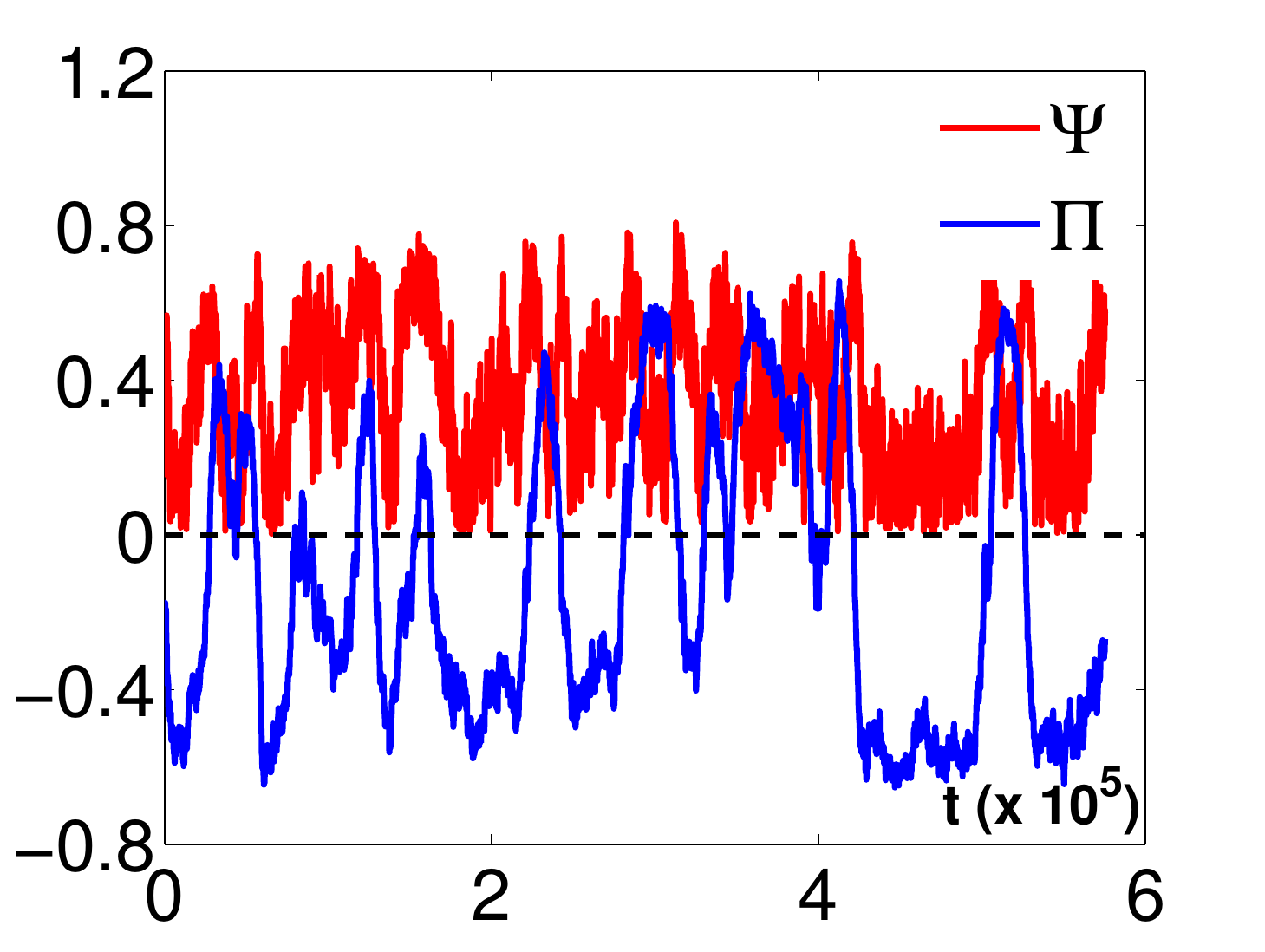}
	\includegraphics[clip,width=8cm,height=5cm]{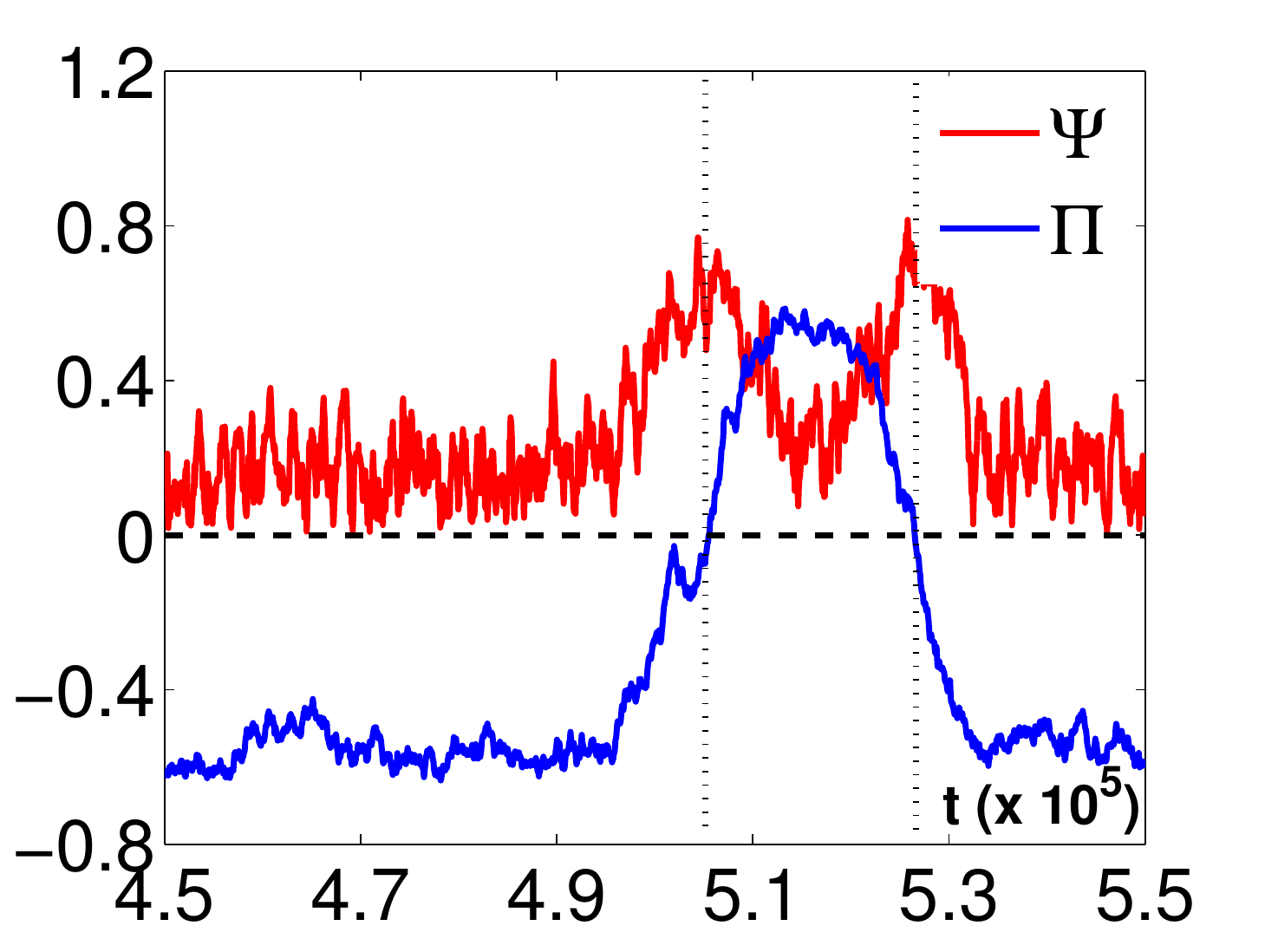}}
	\hbox{\hspace{5cm}(a)\hspace{8cm}(b)}
}
\vspace{3mm}
\vbox{
\hbox{\hspace{0.75cm}
\includegraphics[width=0.30\textwidth]{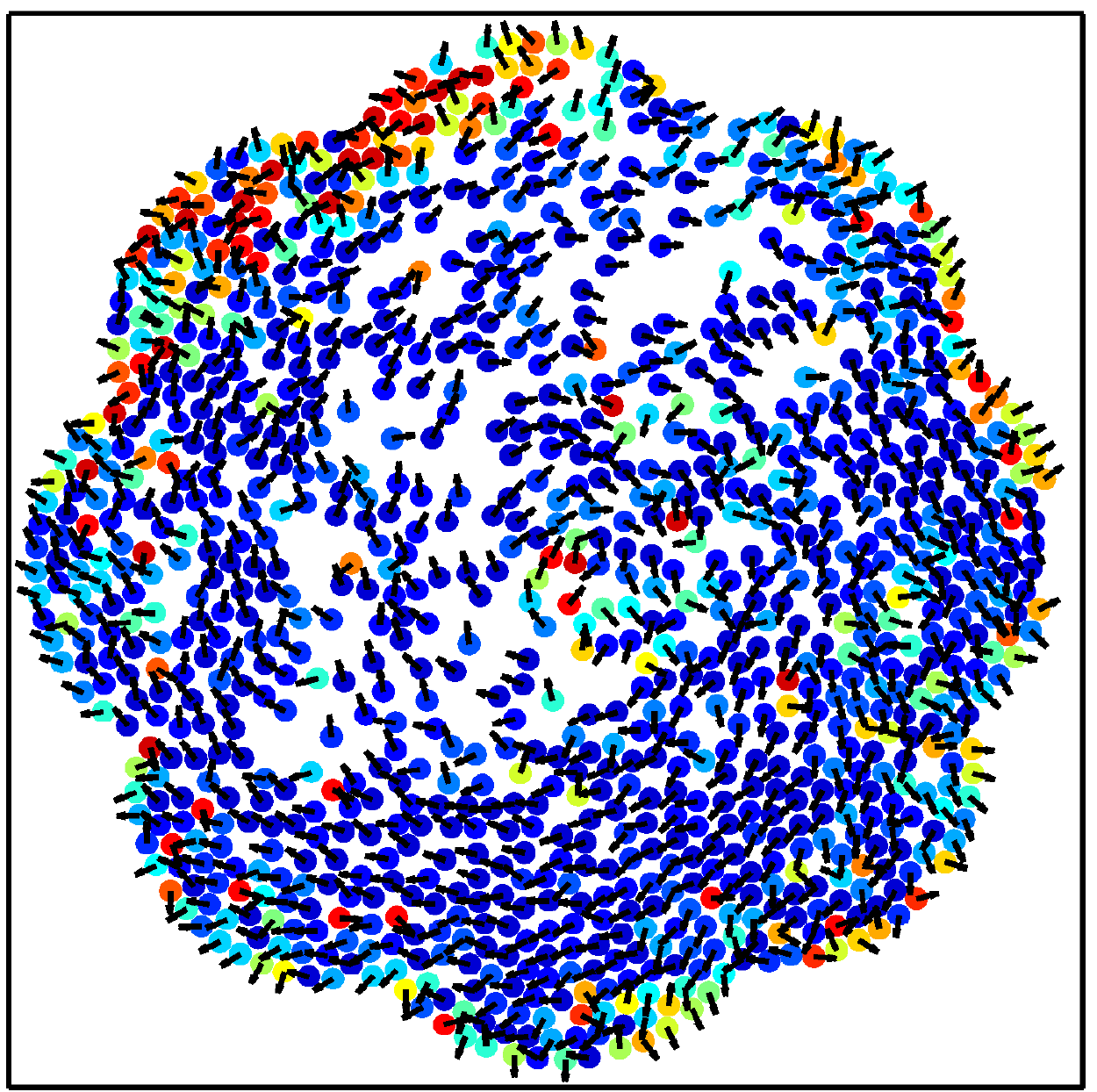}
\includegraphics[width=0.30\textwidth]{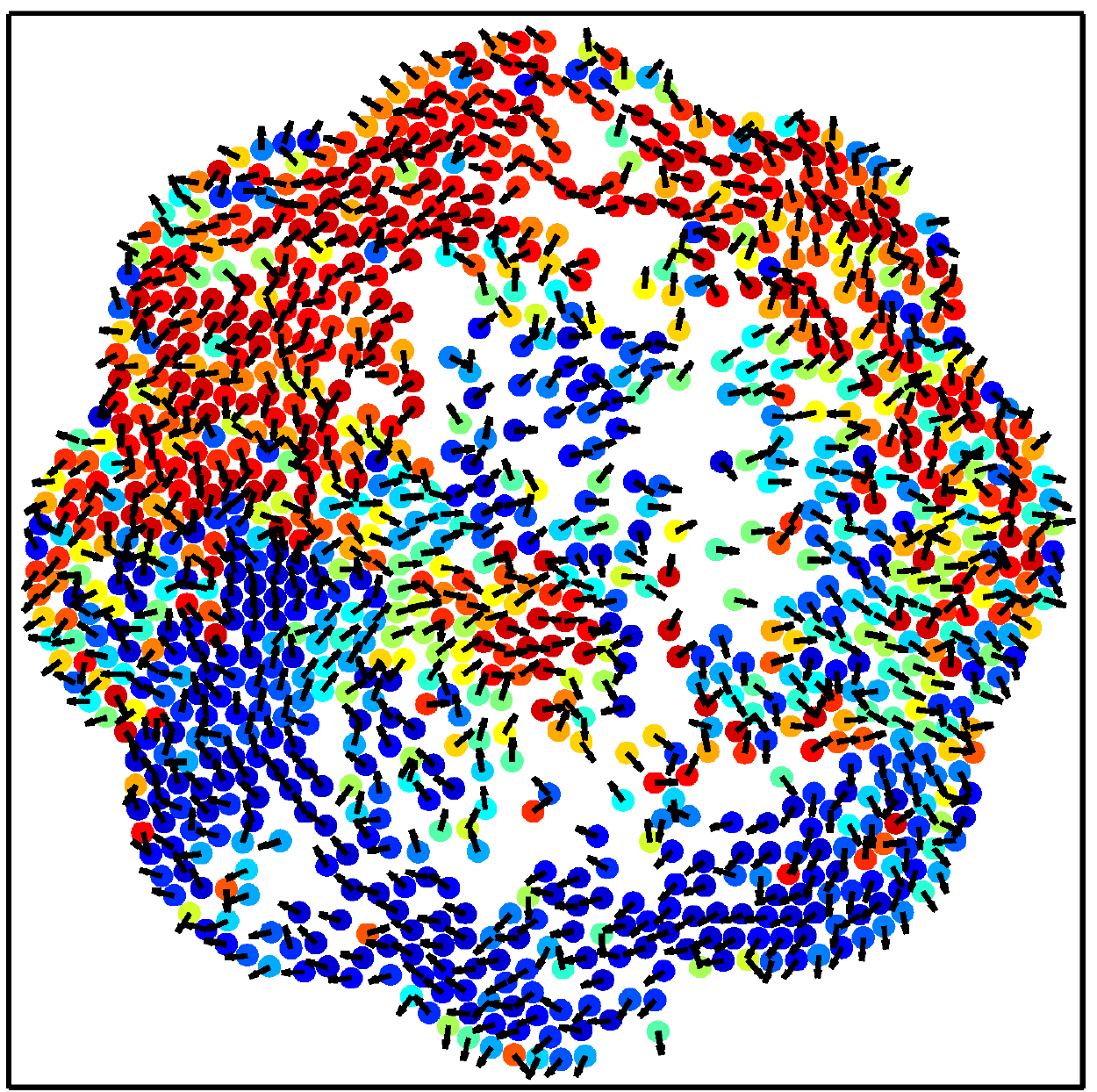}
\includegraphics[width=0.30\textwidth]{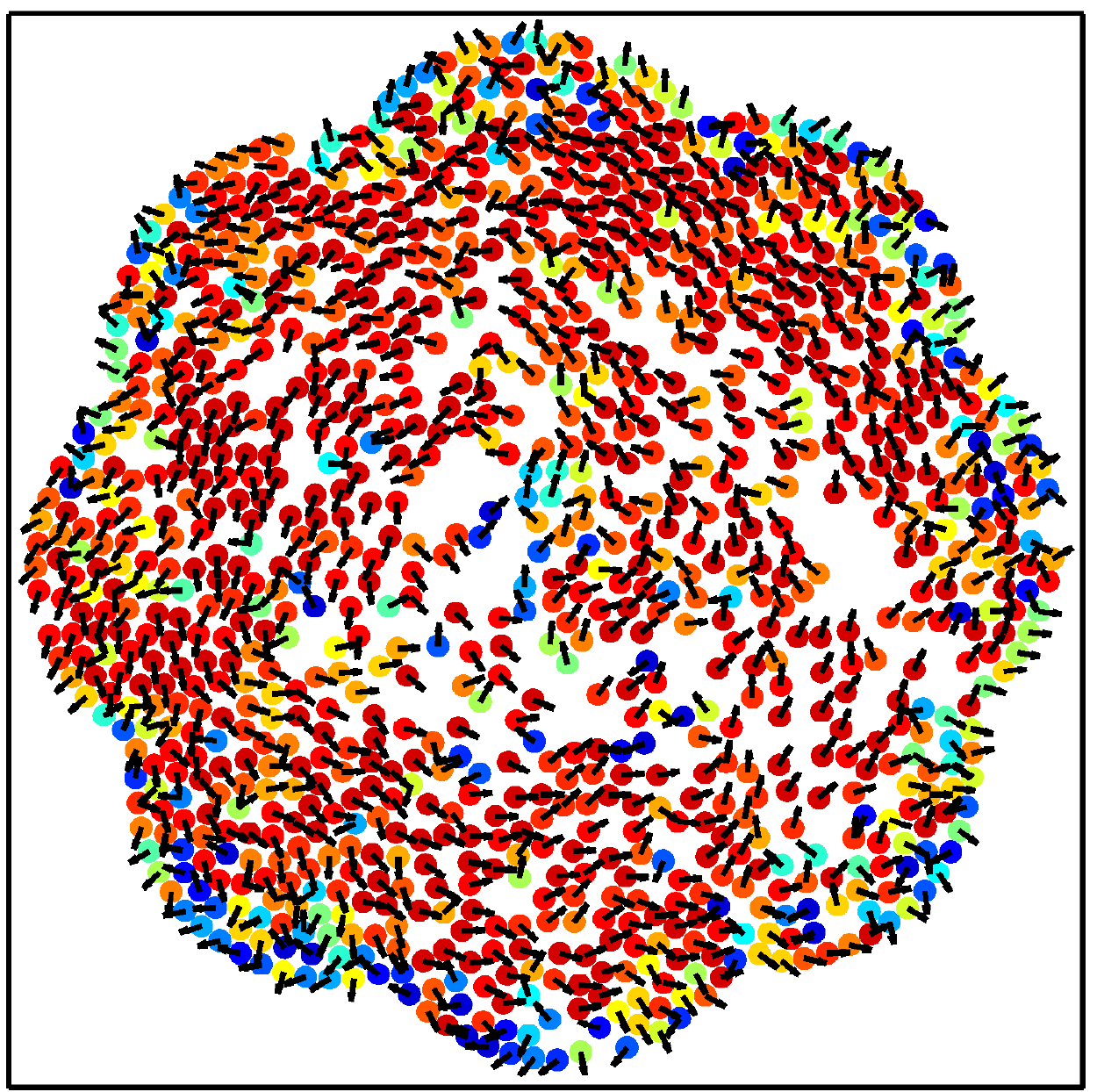} }
\hbox{\hspace{3.5cm}(c)\hspace{5cm}(d)\hspace{5cm}(e)}
}
\caption{(color online) Emergence of confinement-induced milling and its relation to polar order. 
(a) Temporal evolution of the ortho-radial ordering parameter $\Pi(t)$ and the polar ordering parameter $\Psi(t)$. 
(b) Zoom on a time window during which two inversion processes occur (vertical dashed lines) 
(c,d,e) Successive snapshots of the system with, from left to right, the inversion process from a clockwise ($\langle \Pi\rangle <0$ ) to an anti-clockwise $\langle \Pi \rangle >0$ large scale vortex. The packing fraction $\phi=0.58$. 
} 
\label{fig:vortex}
\end{figure*}

We measured number fluctuations in both experiments {\sf Ea} and  {\sf Eb}
for $\Gamma=2.8$. Given the variations of local average packing fraction
described above, we used for $n$ the mean number of particles actually recorded 
in each box in which particles were counted. This actually allowed us to use
boxes slightly outside the ROI for better statistics without any 
detectable problem. As expected, $\Delta n^2$ varies algebraically with $n$ until 
it levels off at large $n$ due to finite-size effects
(Fig.~\ref{fig:finite-size}c). The effective scaling exponent, measured over
two decades, is $\gamma=1.40\pm 0.01$ for experiment {\sf Ea} and  
 $\gamma=1.44\pm 0.01$ for experiment {\sf Eb}. This is close and below the 
expected theoretical value 1.6, in agreement with the observation that the 
asymptotic value is approached from below as the system 
size increases\cite{chate2006simple}.



\subsection{Intermittent confinement-induced milling}

Experiment {\sf Ec} illustrates some of the consequences of increasing 
the density of particles. With a nominal packing fraction $\phi=0.58$,
the ROI being fixed as previously to a diameter of 20, 
we observe an effective packing fraction inside the ROI $\phi_{{\rm ROI}}=0.19$ suggesting
a stronger concentration of the particles along the boundaries, which cannot be ``ignored'' anymore:
the time series of $\Psi$ in the ordered state observed at $\Gamma=3$
show  an intermittent behavior between values 
of the order of 0.6 and values in the range $[0,0.3]$ (Fig.~\ref{fig:vortex}a-b).
These low values do {\it not} indicate a disordered state. 
Rather, they are the trace of a {\it milling} configuration, i.e.
a system-wide vortex obviously resulting from the domain shape
 (Fig.~\ref{fig:vortex}c-e and Supplementary movie \cite{epaps4}).
This can be quantified by calculating the macroscopic angular momentum
$\Pi(t) = \langle\vec{n}_i(t)\dot\vec{e}_{\theta_i}(t)\rangle_i$, 
where $\vec{e}_{\theta_i}(t)$ is the azimuthal unit vector associated to $\theta_i$, 
the angular position of particle $i$ in polar coordinates.
Time series of $\Psi$ and $\Pi$ reveal that  the system switches
intermittently between clockwise and counterclockwise milling: 
$\Pi$ changes sign from time to time, but never dwells long around zero. 
The usual order parameter $\Psi$, in fact, takes large values mostly during these
reversal events (Fig.~\ref{fig:vortex}a-b). 

The transition scenario we have observed in experiments {\sf Ea} and {\sf Eb}
and described  in the previous section is thus strongly modified.
Here when $\Gamma$ is decreased, one first observed the onset of collective
motion, as in the previous case, but soon they organize in system size vortices and the transition
towards polar order is interrupted. Further investigations are required to investigate the interplay
between these two transitions. But one can already conclude that confined active
flows are like any confined flows: boundaries cannot be ignored, and it is hard to
disentangle the intrinsic ``bulk''  properties of active fluids from those resulting from
the inevitable presence of boundaries.

\section{Summary and discussion}
\label{discuss}

Our results constitute, to our knowledge, the first study of a well-controlled 
experimental system in which ``self-propelled'' objects are able to move 
collectively over scales as large as the system size. The dynamics of our particles,
although nominally three-dimensional and complicated, is well accounted for by
a two-dimensional description in terms of persistent random walks. Their binary
collisions are not simple inelastic aligning ones, but are 
spatiotemporally extended events during which multiple actual collisions happen,
leading eventually to a weak but clear effective alignment.

At the collective level, we showed how to avoid the accumulation of particles near the 
boundary walls by adopting flower-shaped arenas whose petals help reinject particles 
in the bulk. This trick is of course not perfect, and we do observe higher average densities
near the walls, so that take must be taken in defining a region of interest in which
the density is nearly constant although different from the nominal packing fraction.
We presented the results of three sets of experiments conducted near ``optimal'' densities 
such that well-developed collective motion is observed, in which orientational order
is established on scales comparable to the system size. In these most ordered regimes,
we recorded clear, unambiguous evidence of ``giant number fluctuations'', a 
signature feature of orientationally-ordered phases in active matter.

From the wealth of numerical studies of models of active matter,
where periodic boundary conditions are often adopted,
it is easy to forget about the inevitable role played by walls
when investigating collective motion. Here, in addition to the subtle effects about
the effective packing fraction recorded in the region of interest, 
we showed that at densities slightly larger than those used in the regimes where
giant number fluctuations were recorded, the boundaries ``come back'' in the problem
by driving macroscopic vortical flows.

The experiments we conducted have also shown some clear limitations of our setup:
Even though we used two domain sizes, finite-size effects remain 
strong and difficult to estimate, which rules out any statement about the (asymptotic) nature of the
transition to collective motion.
Also, because our particles cease to move below some vibration amplitude value, 
we could not explore regimes ``deep in the ordered phase''. 
Thus, to be true, it is hard to disentangle, in the strong density fluctuations
measured, those intrinsic to the ordered phase from
those linked to the proximity of the transition (not to mention again finite-size effects).
Moreover, recent theoretical progress has shown that the ordered 
phase, in a parameter region bordered by the transition, is generically endowed with the
emergence of high-density high-order bands \cite{chate2008collective,bertin2006boltzmann}. 
Here, we did not record any evidence of such bands,
but one can argue that this is only because our system is too small. If these bands exist,
then there is no theoretical reason to believe that the fluctuations exponent $\gamma=8/5$
predicted by Toner and Tu should be found, as their result applies to the (fluctuating)
homogeneous phase typically observed deep enough on the ordered size of the transition,
but not when bands are present.

The above remarks stress the difficulties inherent to systems of vibrated granular 
particles. In order to approach asymptotic regimes closer, one should
use larger domains and/or smaller particles, but both are technically difficult. 
Reverting to the study of large swarms of small robots 
\cite{bobadilla2011manipulating}, or to biological motility assays \cite{schaller467polar},
would suppress limitations in domain size, but at the price of some lesser control.
Another approach, that we are currently pursuing \cite{weber2012numerics}, 
is to study {\it in silico} the system sizes and boundary conditions
experimentally inaccessible by constructing a numerical model faithful
to every experimental finding. Obviously, such a model
cannot be yet another Vicsek-style model. It has to incorporate ingredients such
as the treatment of the polarity vs. velocity dynamics, and to account for
the complex series of collisions typically observed when two particles meet.
It is our hope that such a model could give us access to the truly large-scale behavior
of our system, and maybe, unveil new asymptotic properties.
 
We thank M. van Hecke for advice in the design of our system,  V. Padilla and C. Gasquet
for technical assistance, and E. Bertin for enlightening discussions. 
This work was supported by the French ANR project DyCoAct.


\end{document}